\documentclass[11pt]{article}
\pdfoutput=1
\usepackage{graphicx}
\usepackage{amssymb}
\usepackage{epstopdf}
\usepackage{epsfig}
\usepackage[affil-it]{authblk}
\def\ra{\!\rightarrow\!}

\def\dbar{\overline{D}{}^{\,0}}

\def\dkkpp{$D^0\ra K^+K^-/\pi^+\pi^-$}
\def\cp{$CP$}
\def\cpv{$CPV$}

\def\hb{{\it HERA-B\/}}
\def\dkpi{$D^0\ra K^+\pi^-$}

\def\gevm{~GeV/$c^2$}

\def\belle{Belle}
\def\babar{BaBar}
\def\fbinv{~fb$^{-1}$}
\def\simge{\mathrel{%
   \rlap{\raise 0.511ex \hbox{$>$}}{\lower 0.511ex \hbox{$\sim$}}}}
\def\simle{\mathrel{/?
   \rlap{\raise 0.511ex \hbox{$<$}}{\lower 0.511ex \hbox{$\sim$}}}}

\newcommand{\AmS}{{\protect\the\textfont2
  A\kern-.1667em\lower.5ex\hbox{M}\kern-.125emS}}
\newcommand{\numu}

\title{Renaissance of the $\sim$1~TeV Fixed-Target Program}
\begin{document}
\author[9]{T.~Adams}
\author[7]{J.~A.~Appel}
\author[12]{K.~E.~Arms}
\author[21]{A.~B.~Balantekin}
\author[10]{J.~M.~Conrad}
\author[7]{P.~S.~Cooper} 
\author[6]{Z.~Djurcic}
\author[18]{W.~Dunwoodie}
\author[3]{J.~Engelfried}
\author[10]{P.~H.~Fisher}
\author[7]{E.~Gottschalk}
\author[14]{A.~de Gouvea}
\author[12]{K.~Heller}
\author[10]{C.~M.~Ignarra}
\author[10]{G.~Karagiorgi}
\author[7]{S.~Kwan}
\author[1]{W.~A.~Loinaz}
\author[5]{B.~Meadows}
\author[7]{R.~Moore}
\author[7]{J.~G.~Morf\'{\i}n}
\author[16]{D.~Naples}
\author[17]{P.~Nienaber}
\author[13]{S.~F.~Pate}
\author[13]{V.~Papavassiliou}
\author[11,20]{A.~A.~Petrov}
\author[19]{M.~V.~Purohit}
\author[8]{H.~Ray}
\author[4]{J.~Russ}
\author[5]{A.~J.~Schwartz}
\author[6]{W.~G.~Seligman}
\author[6]{M.~H.~Shaevitz}
\author[14]{H.~Schellman}
\author[,22]{\indent J.~Spitz\footnote{Corresponding author\newline \indent \textit{Email address: joshua.spitz@yale.edu} (J. Spitz)}}
\author[7]{M.~J.~Syphers}
\author[2,14]{T.~M.~P.~Tait}
\author[15]{F.~Vannucci}
\affil[1]{Amherst College, Amherst, MA 01002} 
\affil[2]{Argonne National Laboratory, Argonne, IL 60439} 
\affil[3]{Universidad Aut\'onoma de San Luis Potos\'{\i}, Mexico 78240}
\affil[4]{Carnegie-Mellon University, Pittsburgh, PA 15213}
\affil[5]{University of Cincinnati, Cincinnati, OH 45221}
\affil[6]{Columbia University, New York, NY 10027}
\affil[7]{Fermi National Accelerator Laboratory, Batavia, IL 60510}
\affil[8]{University of Florida, Gainesville, FL 32611}
\affil[9]{Florida State University, Tallahassee, FL 32306}
\affil[10]{Massachusetts Institute of Technology, Cambridge, MA 02139}
\affil[11]{University of Michigan, Ann Arbor, MI 48201}
\affil[12]{University of Minnesota, Minneapolis, MN 55455}
\affil[13]{New Mexico State University, Las Cruces, NM 88003} 
\affil[14]{Northwestern University, Chicago, IL 60208}
\affil[15]{University Paris 7, APC, Paris, France}
\affil[16]{University of Pittsburgh, Pittsburgh, PA 15260}
\affil[17]{Saint Mary's University of Minnesota, Winona, MN 55987}
\affil[18]{SLAC National Accelerator Laboratory, Stanford, CA 94309}
\affil[19]{University of South Carolina, Columbia, SC 29208}
\affil[20]{Wayne State University, Detroit, MI 48201}
\affil[21]{University of Wisconsin, Madison, WI 53706} 
\affil[22]{Yale University, New Haven, CT 06520}

\maketitle

\begin{abstract}
This document describes the physics potential of a new
fixed-target program based on a $\sim$1~TeV proton source.  Two
proton sources are potentially available in the future: the existing
Tevatron at Fermilab, which can provide 800~GeV protons for fixed-target physics, and a possible upgrade to the SPS at CERN, called
SPS+, which would produce 1~TeV protons on target. In this paper we
use an example Tevatron fixed-target program to illustrate the high
discovery potential possible in the charm and neutrino sectors.  We
highlight examples which are either unique to the program or
difficult to accomplish at other venues.
\end{abstract}

\section{Introduction}

{\it Fixed-target at approximately TeV energies?  Didn't we do that
  for over twenty years ending a decade ago?  Why revisit that
  strategy?}

A renaissance in TeV-energy fixed-target physics has become possible
because of new detector technologies and improvements in accelerators
since the 1990's.  As a result, we can describe a fixed-target physics program, focusing on the charm and neutrino sectors.
The program is unique to a $\sim$1~TeV fixed-target facility
and complements the ongoing physics program envisioned by the
community for the late 2010's.

There are two possible sources of $\sim$1~TeV protons which may be
available.  The first is the Tevatron at Fermilab, which can be
modified for fixed-target running.  Details on how this machine can be
run at higher intensity and higher efficiency than in the past are
discussed in Appendix A of this paper.  The second possible source is
the SPS+~\cite{sps,sps2} which is planned at CERN as part of the LHC upgrade program.
The fixed-target program described here can run during times when the SPS+
is not providing beam to LHC.  The energy of SPS+ is expected to be
about 1~TeV.  For the results presented here, we have assumed 800~GeV
protons on target since this is the capability of the existing
machine. However, the physics case only improves for running at 1~TeV.

The purpose of this paper is to illustrate the strength and richness
of an envisioned fixed-target program.  In particular, this paper concentrates
on a new study of discovery potential in the charm sector, which would
utilize slow-spill beams.  A future $D^0$-$\dbar$ mixing and \cp\
violation (\cpv) experiment with three years of running could reconstruct an
order of magnitude more flavor-tagged \dkpi\ decays than will be
reconstructed by the $B$-factory experiments with their full data
sets. The resulting sensitivity to \cpv\ parameters $|q/p|$
and ${\rm Arg}(q/p)$ is found to be much greater than current world
sensitivity.  However, to illustrate that this is a well-rounded
program, we also explore ideas in the neutrino sector.  We review the
case for a precision electroweak neutrino experiment running from a
very pure sign-selected high energy $\nu_\mu$ beam, which has been
discussed in more detail elsewhere~\cite{NuSOnGEW,NuSOnGQCD} and we
present new studies on two promising and unique avenues for beyond
Standard Model neutrino searches using beam dump production.  The
first of these uses $\nu_\tau$ charged current events 
produced by a proton beam in the 800~GeV to 1~TeV
range.  The second is a search for neutral heavy leptons produced in
the beam dump. Emphasizing the breadth of physics possible in the high-energy neutrino scattering sector of this new fixed target program, an extensive study of high-precision QCD is described in a separate paper~\cite{NuSOnGQCD}.

This combination of experiments represents an integrated program aimed
at discovery of new physics.  At the same time, each of these
experiments will provide a wide array of interesting and valuable
measurements within the Standard Model.  The program is very
physics rich and will provide opportunities for many physicists.  The result is a
compelling opportunity for the future.

\section{The Discovery Potential of Fixed-Target Charm}

\subsection{Introduction}

As mentioned, there was a very successful fixed-target charm 
program at the Fermilab Tevatron~\cite{fnal_fixed-target}. 
Not only did it provide high precision measurements (some of which remain 
the most precise even today), but it also advanced flavor physics thinking 
in a way that still underlies many current analyses. It also demonstrated 
the utility of precision vertexing for heavy flavor physics, paving the way 
for the incorporation of silicon tracking systems in all the latest experiments.  
The fixed-target charm program ended when the applied technologies were 
more-or-less played out, and attention turned to the opportunities at 
colliders, both at $e^+e^-$ and hadron machines.  The reason to now revisit
the possibility of a fixed-target charm experiment is a combination 
of increased interest in charm mixing (now observed) and possible \cpv\ in the charm system, and the availability 
of technology well beyond what was available at the end of the previous 
program~\cite{Schwartz,Buras,Grossman1, Golowich,Grossman2,Li,Ball,Blanke,Nir}. A Tevatron fixed-target experiment may be the most cost-effective 
way forward in the charm sector, as the Tevatron would not 
need to be run in collider mode. Also, the beam energy could 
be reduced and still remain far above charm production threshold.
Such an experiment at the Tevatron has the potential to greatly improve upon 
the sensitivity to mixing and \cpv\ achieved by the B factories. 
We note that the most 
sensitive measurements of mixing and \cpv\ rely on measuring decay-time 
distributions. For this type of measurement, a fixed-target experiment 
has an advantage over an $e^+e^-$ $B$ factory experiment due to the fact that 
the mean decay length is notably larger than the vertex resolution.
We will address these physics opportunities below.
In the recent ``Roadmap for US High-Energy Physics'' written by the 
Particle Physics Project Prioritization Panel (P5), future operation of 
the Tevatron was not considered. 
However, there exists a plan to keep the Tevatron cold after completion 
of the collider program such that it could easily be operated again should 
sufficiently compelling physics opportunities arise. 

   
\vskip0.30in

A Fermilab Tevatron fixed-target experiment could
produce very large samples of $D^*$ mesons that decay via 
$D^{*+}\ra D^0\pi^+,\ D^0\ra K^+\pi^-$~\footnote{Charge-conjugate modes 
are implicitly included unless noted otherwise.}. 
The decay time distribution of the ``wrong-sign'' $D^0\ra K^+\pi^-$ 
decay is sensitive to $D^0$-$\dbar$ mixing parameters $x$ and~$y$. 
Additionally, comparing the $D^0$ decay time distribution to that 
for $\dbar$ allows one to measure or constrain the \cpv\ parameters $|q/p|$ and ${\rm Arg}(q/p)\equiv\phi$. This 
method has been used previously by Fermilab experiments  
E791~\cite{e791_kpi} and E831~\cite{e831_kpi} to search 
for $D^0$-$\dbar$ mixing. However, those 
experiments ran in the 1990's and reconstructed only a few hundred 
flavor-tagged $D^0\ra K^+\pi^-$ decays. Technological advances in 
vertexing detectors and electronics made since then make a much improved fixed-target experiment possible. We 
estimate the expected sensitivity of such an experiment, and compare 
it to that of the $B$ factory experiments \belle\ and \babar. 
Those experiments have reconstructed several thousand signal decays 
and using these samples, along with those for \dkkpp, have made the 
first observation of $D^0$-$\dbar$ mixing~\cite{babar_kpi,belle_kk}.
The CDF experiment has also measured 
$D^0$-$\dbar$ mixing using $D^0\ra K^+\pi^-$ decays~\cite{cdfD0mix}. 
Although the background is much higher than at an $e^+e^-$ experiment,
the number of reconstructed signal decays is larger, and the statistical 
errors on the mixing parameters are similar to those of \babar.

Although we focus on measuring $x,~y,~|q/p|$, and $\phi$, a much 
broader charm physics program is possible at a Tevatron experiment.
We also briefly present some of these other opportunities.

\subsection{Expected signal yield}

We estimate the expected signal yield by scaling from two previous
fixed-target experiments, E791 at Fermilab and \hb\ at DESY. These 
experiments had center-of-mass energies and detector geometries 
similar to those that a new charm experiment at the Tevatron would 
have.

\subsubsection{Scaling from \emph{HERA-B}}

\hb\ took data with various trigger configurations. One configuration
used a minimum-bias trigger, and from this data set the experiment
reconstructed $61.3\,\pm 13$ $D^*$-tagged ``right-sign''
$D^0\ra K^-\pi^+$ decays in $182\times 10^6$ hadronic interactions~\cite{herab_kpi}. 
This yield was obtained after all selection requirements were applied. 
Multiplying this rate by the ratio of doubly-Cabibbo-suppressed to 
Cabibbo-favored decays
$R^{}_D\equiv\Gamma(D^0\ra K^+\pi^-)/\Gamma(D^0\ra K^-\pi^+)=0.380\%$~\cite{dcstocf}
gives a rate of reconstructed, tagged \dkpi\ decays per hadronic interaction
of $1.3\times 10^{-9}$. To estimate the sample 
size a Tevatron experiment would reconstruct, we assume the experiment 
could achieve a similar fractional rate. If the experiment ran at an 
interaction rate of 7~MHz (which was achieved by \hb\ using a two-track 
trigger configuration), and took data for $1.4\times 10^7$ live seconds 
per year, then it would nominally reconstruct 
$(7{\rm\ MHz})(1.4\times 10^7)(1.3\times 10^{-9})(0.5)=64000$ flavor-tagged 
\dkpi\ decays per year, or 192000 decays in three years of running. 
Here we have assumed a trigger efficiency of 50\% relative to that 
of~\hb, which is a simple estimate: the trigger needs to be more 
restrictive than the minimum-bias configuration of \hb, but,
on the other hand, the technology has advanced since \hb\ 
was designed and the trigger latency and other inefficiencies 
should be substantially reduced.

\subsubsection{Scaling from E791}

Fermilab E791 was a charm hadroproduction experiment that took data
during the 1991-1992 fixed-target run. The experiment ran with a modest 
transverse-energy threshold trigger, and it reconstructed 35 $D^*$-tagged
\dkpi\ decays in $5\times 10^{10}$ hadronic interactions~\cite{e791_kpi}.
This corresponds to a rate of $7\times 10^{-10}$ reconstructed decays per 
hadronic interaction. Assuming a future Tevatron experiment achieves
this fractional rate, one estimates a signal yield of
$(7{\rm\ MHz})(1.4\times 10^7)(7\times 10^{-10})=69000$ per year, 
or 207000 in three years. This value is similar to that obtained by scaling 
from \hb. We have assumed the same trigger\,+\,reconstruction efficiency 
as that of E791. 
We note that
E791 had an inactive region in the middle of the tracking stations 
where the $\pi^-$ beam passed through, and a future Tevatron 
experiment could avoid this acceptance loss. We do not include 
any improvement for this in our projection.

\subsection{Comparison with the $B$ factories}

We compare these yields with those that will
be attained by the $B$ factory experiments after they have
analyzed all their data. The \belle\ experiment reconstructed 4024
$D^*$-tagged \dkpi\ decays in 400\fbinv\ of data~\cite{belle_kpi},
and it is expected to record a total of 1000\fbinv\ when it completes
running. This integrated luminosity corresponds to 10060 signal events. 

The \babar\ experiment reconstructed 4030 tagged \dkpi\ decays 
in 384\fbinv\ of data~\cite{babar_kpi}, and the experiment recorded 
a total of 484\fbinv\ when it completed running in early~2008. 
Thus the total \babar\ data set corresponds to 5080 signal 
events. Adding this to the estimated final yield from \belle\ 
gives a total of 15100 \dkpi\ decays. This is less than 8\% of 
the yield estimated for a Tevatron experiment in three years 
of running.

The KEK-B accelerator where \belle\ runs is scheduled to be 
upgraded to a ``Super-$B$'' factory running at a luminosity 
of $\sim\!8\times 10^{35}$~cm$^{-2}$\,s$^{-1}$~\cite{superbelle}. There is
also a proposal to construct a Super-$B$ factory in Italy near 
the I.N.F.N. Frascati laboratory~\cite{superB}. An experiment at 
either of these facilities would reconstruct very large 
samples of $D^{*+}\ra D^0\pi^+,\,D^0\ra K^+\pi^-$ decays. In fact 
the resulting sensitivity to $x'^2$ and $y'$ may be dominated 
by systematic uncertainties. This merits further study. We 
note that many of the systematic errors obtained at a future
Tevatron experiment are expected to be smaller than those
at an $e^+e^-$ collider experiment, due to the superior
vertex resolution and $\pi/K$ identification possible 
with a forward-geometry detector.

\subsection{Comparison with hadron colliders}

The LHCb experiment has a forward geometry
and is expected to reconstruct $D^{*+}\ra D^0\pi^+,\,D^0\ra K^+\pi^-$ 
decays in which the $D^*$ originates from a $B$ decay.
The resulting sensitivity to mixing 
parameters $x'^2$ and $y'$ has been studied in Ref.~\cite{lhcb_kpi}.
This study assumes a $b\bar{b}$ cross section of 500~$\mu$b and
estimates several unknown trigger and reconstruction efficiencies.
It concludes that approximately 58000 signal decays would be reconstructed
in 2\fbinv\ of data, which corresponds to one year of running. This 
yield is similar to that estimated for a Tevatron experiment. 
However, LHCb's trigger is efficient only for $D$ mesons having
high $p^{}_T$, {\it i.e.} those produced from $B$ decays. This
introduces two complications:
\begin{enumerate}
\item Some fraction of prompt $\dbar\ra K^+\pi^-$ decays will be 
mis-reconstructed or undergo multiple scattering and, after being 
paired with a random soft pion, will end up in 
the \dkpi\ sample (fitted for $x'^2$ and $y'$). As the 
production rate of prompt $D$'s is $~$two orders of magnitude
larger than that of $B$'s, this component may be non-negligible,
and thus would need to be well-understood when fitting.
\item To obtain the $D^*$ vertex position ({\it i.e.} the
origin point of the $D^0$), the experiment must reconstruct
a $B\ra D^* X$ vertex, and the efficiency for this is not 
known. Monte Carlo studies indicate it is 51\%~\cite{lhcb_kpi}, 
but there is uncertainty in this value.
\end{enumerate}

The LHCb study found that, for $N^{}_{K^+\pi^-}=232500$,
a signal-to-background ratio ($S/B$) of 0.40, and a decay
time resolution ($\sigma^{}_t$) of 75~fs, the statistical 
errors obtained for $x'^2$ and $y'$ were 
$6.4\times 10^{-5}$ and $0.87\times 10^{-3}$, respectively. 
These values are less than half of those that we estimate 
can be attained by the $B$ factories by scaling current 
errors by $\sqrt{N_{K^+\pi^-}}$:
$\delta x'^2\approx 14\times 10^{-5}$ and 
$\delta y'\approx 2.2\times 10^{-3}$.
As the signal yield, $S/B$, and $\sigma^{}_t$ of a future Tevatron 
experiment are similar to those for LHCb, we expect that similar 
errors for $x'^2$ and $y'$ can be attained.

The CDF measurement of charm mixing~\cite{cdfD0mix} uses 12700 
$D^{*+}\ra D^0\pi^+,\,D^0\ra K^+\pi^-$ decays from 1.5\fbinv\ of
integrated luminosity.  This could increase by about a factor 
of five by the end of Run II at the Tevatron collider. 
Assuming that both statistical and systematic errors are reduced 
by the square root of the anticipated increase in luminosity, 
one estimates errors of
$16\times 10^{-5}$ and $3.4\times 10^{-3}$ for
$\delta x'^{2}$ and $\delta y'$, respectively,
at the end of Run II. 

\begin{figure}[pb]
\centering
\begin{tabular}{cc}
\includegraphics[width=2.4in]{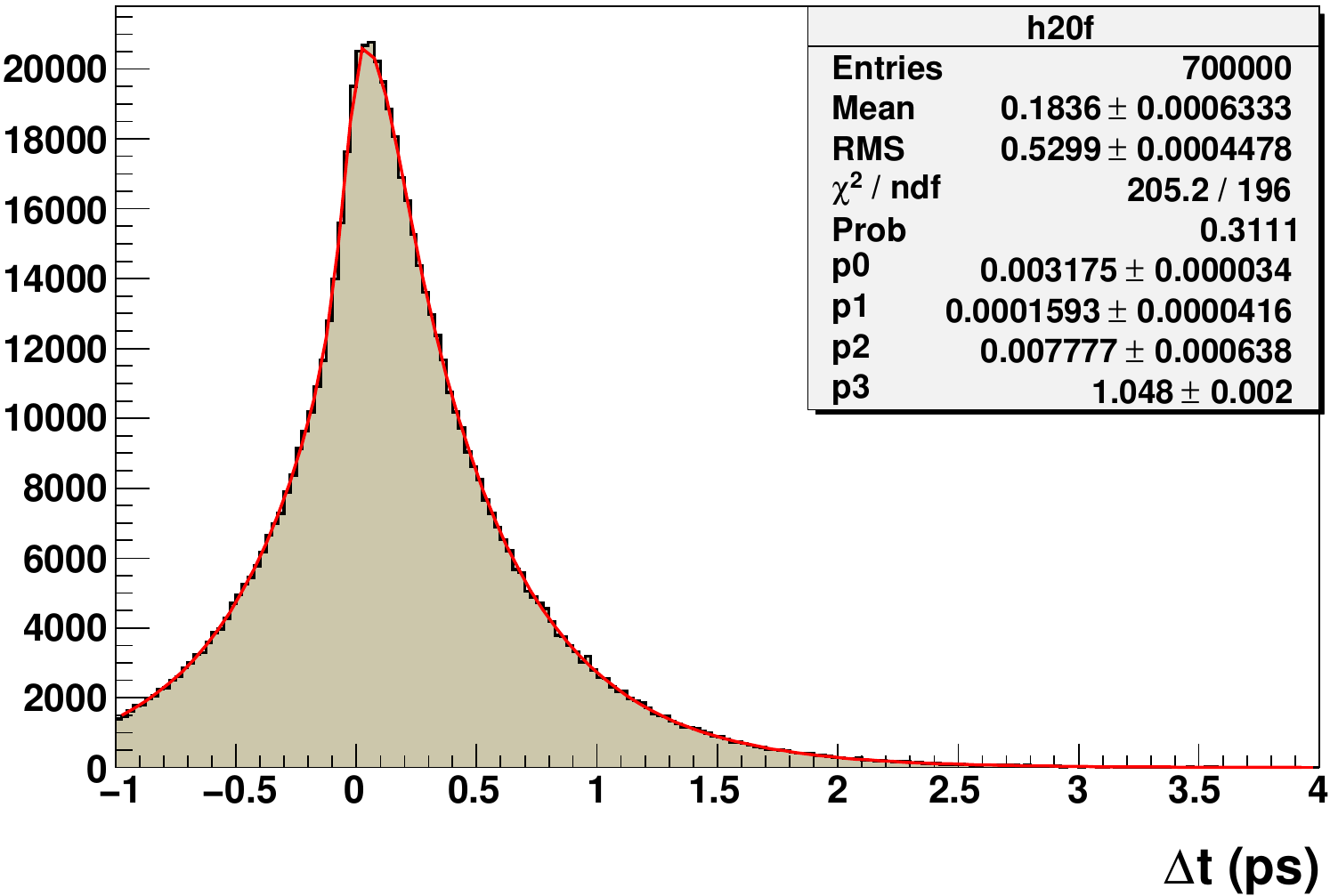}
\hskip0.3in
\includegraphics[width=2.4in]{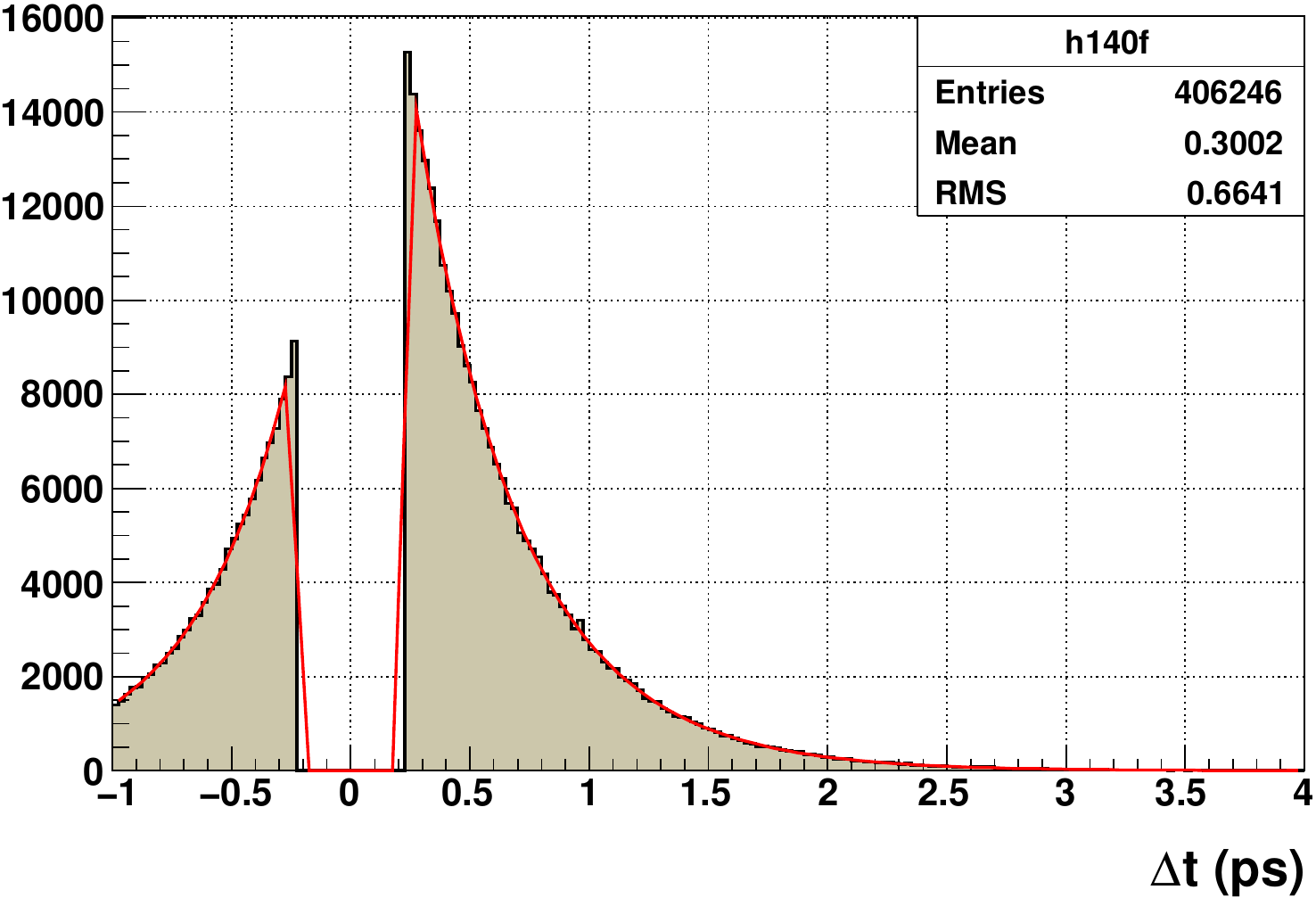}
\end{tabular}
\caption{Monte Carlo \dkpi\ decay time distributions (left) 
without and (right) with a minimum decay time cut.
Superimposed is the result of a fit. The ratio of signal to
background after the $t^{}_{\rm min}$ ($=\!\tau^{}_D/2$) cut is 
0.40, and the decay time resolution $\sigma^{}_t$ is 75~fs.}
\label{fig:toymc_fit}
\end{figure}

To compare to 
these estimates, we have done a ``toy'' Monte Carlo (MC) 
study to estimate the sensitivity of a Tevatron experiment.
The results obtained are similar to those of LHCb:
for $N^{}_{K^+\pi^-}=200000$, $S/B=0.40$, $\sigma^{}_t=75$~fs,
and a minimum decay time cut of $0.5\times\tau^{}_D$
(to reduce combinatorial background), we find 
$\delta x'^2=5.8\times 10^{-5}$ and 
$\delta y'=1.0\times 10^{-3}$.
These errors are the RMS's of the distributions of
residuals obtained from fitting an ensemble of 200 experiments.
A typical fit is shown in Fig.~\ref{fig:toymc_fit}. 

Note that it is difficult to know when a Tevatron charm experiment 
might be performed and results available. That makes it challenging 
to say what may be the world situation by the time such an experiment 
is done. The point of this paper is to say what such an experiment 
might achieve.

\subsection{Global fit for \cpv\ parameters}

If we assume the $\delta x'^2$ and $\delta y'$ errors
obtained in our toy MC study (which are close to the values
obtained in the LHCb study), we can estimate the resulting 
sensitivity to \cpv\ parameters $|q/p|$ and~$\phi$.
The first parameter characterizes \cpv\ in the mixing
of $D^0$ and $\dbar$ mesons, while the second parameter
is a phase that characterizes \cpv\ resulting from interference
between an amplitude with mixing and a direct decay amplitude.
In the Standard Model, $|q/p|$ and $\phi$ are essentially 
1 and 0, respectively. A measurable deviation from these 
values would indicate new physics.

To calculate the sensitivity to $|q/p|$ and $\phi$,
we do a global fit of eight underlying parameters to 28 
measured observables. The fitted parameters are $x$ and $y$,
strong phases $\delta^{}_{K\pi}$ and $\delta^{}_{K\pi\pi}$, $R^{}_D$,
and \cpv\ parameters $A^{}_D,\,|q/p|$ and~$\phi$.
Our fit is analogous to that done by the Heavy Flavor
Averaging Group (HFAG)~\cite{hfag_charm_fits}. The only
difference is that we reduce the errors for $x'^2$ and $y'$
according to our toy MC study, and we also reduce the error 
for $y^{}_{CP}$ by a similar fraction. This latter parameter is 
measured by fitting the decay time distribution of \dkkpp\ 
decays, which would also be triggered on and reconstructed 
by a Tevatron charm experiment.

\begin{figure}[h!]
\centering
\begin{tabular}{cc}
\includegraphics[width=2.4in]{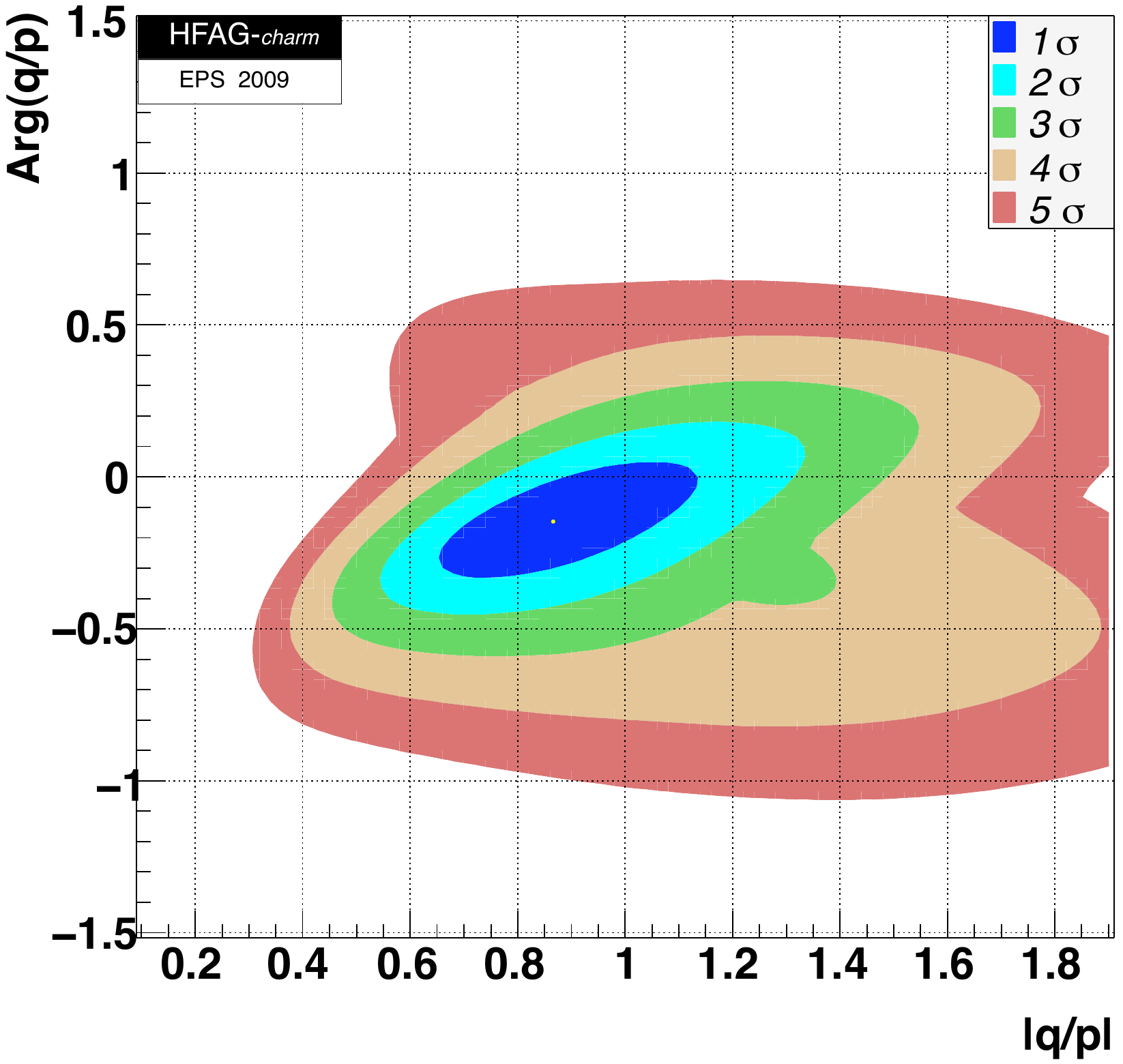}
\hskip0.3in
\includegraphics[width=2.4in]{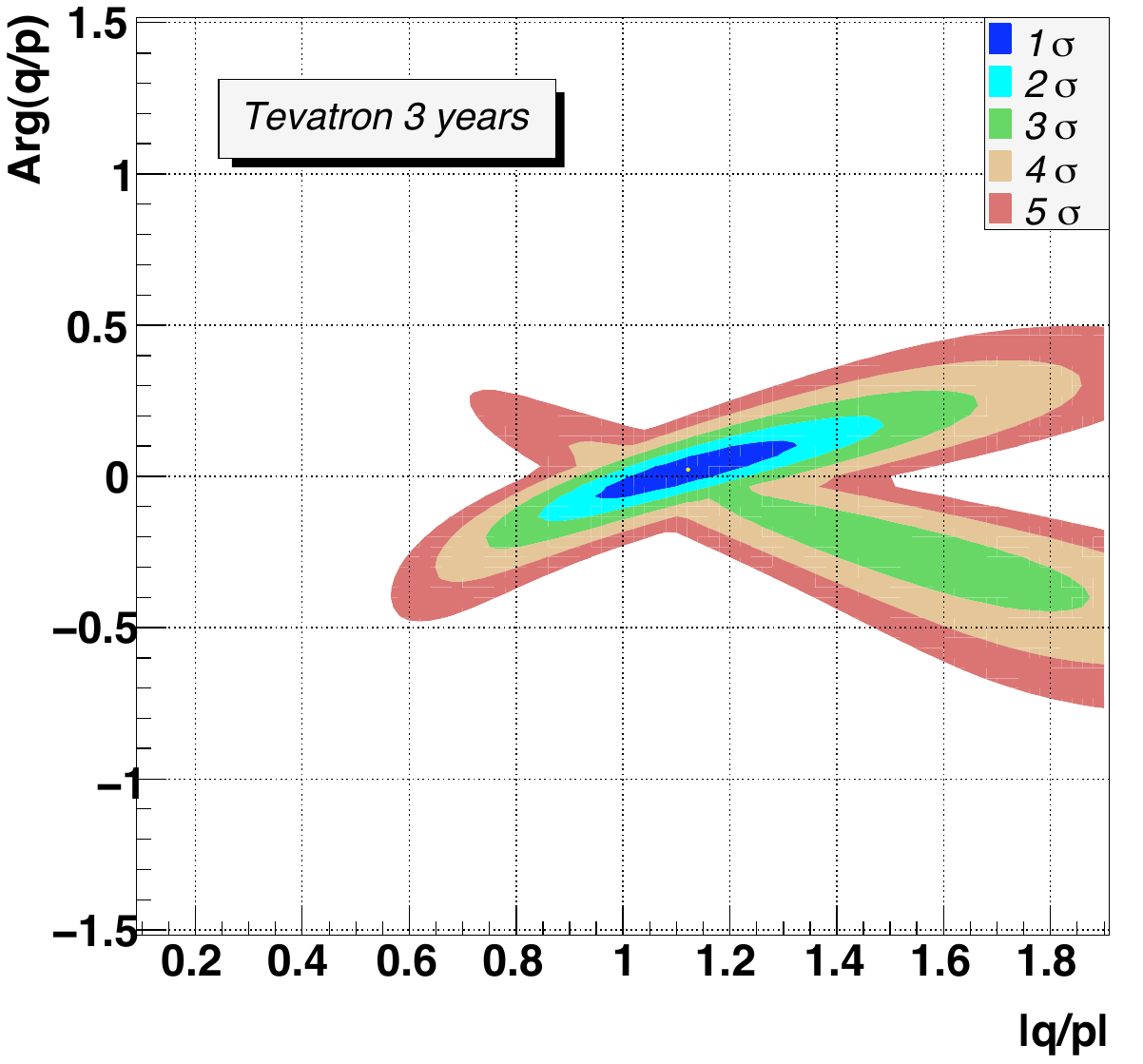}
\end{tabular}
\caption{$|q/p|$ versus $\phi$ likelihood contours resulting 
from a global fit to measured observables (see text).
Left: data after EPS 2009, from HFAG\protect~\cite{hfag_charm_fits}. 
Right: after three years of running of a Tevatron charm experiment.}
\label{fig:fit_results}
\end{figure}

The results of the fit are plotted in Fig.~\ref{fig:fit_results} (right). 
The figure shows two-dimensional likelihood contours for $|q/p|$ 
and~$\phi$. For comparison, the analogous HFAG plot as presented
at the EPS 2009 conference is shown in Fig.~\ref{fig:fit_results} (left). 
One sees that a future Tevatron experiment would yield a very
substantial improvement. 
However, by the time a Tevatron experiment runs, could the 
world situation be different from that shown in the HFAG plot?
Much of the constraining power in the plot is due to the 
measurement of 
$x'$ and $y'$ in $D^0\ra K^+\pi^-$ decays, and of
$y^{}_{CP}$ in $D^0\ra K^+K^-/\pi^+\pi^-$ decays.
For these observables, the Belle/BaBar data sets used 
consist of 400/540~fb$^{-1}$ and 384/384~fb$^{-1}$, respectively;
these together comprise about 65\% of the total data set.
Significant constraining power is also due to measurements of 
$x,\,y,\,|q/p|$, and $\phi$ in $D^0\ra K^0_S\pi^+\pi^-$ decays,
and of $x'',\,y''$ in $D^0\ra K^+\pi^-\pi^0$ decays;
for these measurements only about 30-40\% of the total
data set has been used. Thus we conclude that, once
all Belle and Babar data is analyzed, the errors on
the observables will improve by perhaps a factor 
of~$\sim\!\sqrt{2}$. This is much less than the factor
of 3-4 improvement in these observables used to produce
Fig.~\ref{fig:fit_results} (right).
The addition of more CDF data and of BESIII data will
also improve the HFAG plot, but the improvement on top 
of that due to Belle and Babar is expected to be modest.

\subsection{Other physics}

\subsubsection {Direct \cpv\ searches} 

In addition to searching for \cpv\ arising from
$D^0$-$\dbar$ mixing, one can search for direct \cpv\,
{\it i.e.} \cp\ violation occurring in the decay amplitudes 
themselves. To search for this, one uses $D^+$ or tagged-$D^0$ decays and looks for an asymmetry between the $D\ra f$ and $\overline{D}\ra \bar{f}$ decay rates.

Tables~\ref{tab:cp_charged}-\ref{tab:cp_neutral} are from 
the HFAG~\cite{hfag_charm_dcpv} and 
list current measurements of numerous direct \cpv\ decay
modes. Modes with 2- and 3-body final states could 
potentially be triggered on in a fixed-target
Tevatron experiment and thus could be studied. 
The sensitivity to many of these modes is likely to be
substantially greater than that
of the current generation of experiments.

\begin{table*}[h]
\begin{center}
{\begin{tabular}{|l|c|c|c|} 
\hline
{\bf Mode} & {\bf Year} & {\bf Collaboration} & {\boldmath $A^{}_{CP}$} \\
\hline
{\boldmath $D^+ \to K^0_s\pi^+$} &
  2007 & CLEOc~\cite{CLEOc2007}  &  $ -0.006  \pm 0.010  \pm 0.003 $ \\
&  2002 & FOCUS~\cite{FOCUS2002}  &  $ -0.016  \pm 0.015  \pm 0.009 $ \\
\hline
{\boldmath $D^+ \to K^0_sK^+$} &
  2002 & FOCUS~\cite{FOCUS2002}  &  $ +0.071  \pm 0.061  \pm 0.012 $ \\
\hline
{\boldmath $D^+ \to \pi^+\pi^-\pi^+$} &
  1997 & E791~\cite{E7911997}    &  $ -0.017  \pm 0.042  $ (stat.) \\
\hline
{\boldmath $D^+ \to K^-\pi^+\pi^+$} &
  2007 & CLEOc~\cite{CLEOc2007}  &  $ -0.005  \pm 0.004  \pm 0.009  $ \\
\hline
{\boldmath $D^+ \to K^0_s\pi^+\pi^0$} &
  2007 & CLEO-c~\cite{CLEOc2007} &  $ +0.003  \pm 0.009  \pm 0.003  $ \\
\hline
{\boldmath $D^+ \to K^+K^-\pi^+$} &
  2008 & CLEO-c~\cite{CLEOc2008} &  $ -0.0003  \pm 0.0084  \pm 0.0029  $ \\
&  2007 & CLEO-c~\cite{CLEOc2007} &  $ -0.001  \pm 0.015  \pm 0.008  $ \\
&  2005 & BABAR~\cite{BaBar2005}  &  $ +0.014  \pm 0.010  \pm 0.008 $ \\
&  2000 & FOCUS~\cite{FOCUS2000}  &  $ +0.006  \pm 0.011  \pm 0.005 $ \\
&  1997 & E791~\cite{E7911997}    &  $ -0.014  \pm 0.029  $ (stat.) \\
&  1994 & E687~\cite{E6871994}     &  $ -0.031  \pm 0.068  $ (stat.) \\
\hline
{\boldmath $D^+ \to K^-\pi^+\pi^+\pi^0$} &
  2007 & CLEOc~\cite{CLEOc2007}  &  $ +0.010  \pm 0.009  \pm 0.009  $ \\
\hline
{\boldmath $D^+ \to K^0_s\pi^+\pi^+\pi^-$} &
  2007 & CLEOc~\cite{CLEOc2007}  &  $ +0.001  \pm 0.011  \pm 0.006  $ \\
\hline
{\boldmath $D^+ \to K^0_sK^+\pi^+\pi^-$} &
  2005 & FOCUS~\cite{FOCUS2005}  &  $ -0.042  \pm 0.064  \pm 0.022  $ \\
\hline 
\end{tabular}}
\end{center}
\caption{\cp\ asymmetry 
$A^{}_{CP}= [\Gamma(D^+)-\Gamma(D^-)]/[\Gamma(D^+)+\Gamma(D^-)]$
for $D^\pm$ decays.}
\label{tab:cp_charged}
\end{table*}

\begin{table*}[h]
\begin{center}
{\begin{tabular}{|l|c|c|c|} 
\hline
{\bf Mode} & {\bf Year} & {\bf Collaboration} & {\boldmath $A^{}_{CP}$} \\
\hline
{\boldmath $D^0 \to \pi^+\pi^-$} &
 2008 & Belle~\cite{Belle2008_2} & $ +0.0043 \pm 0.0052 \pm 0.0012 $ \\
& 2008 & BABAR~\cite{BaBar2008} & $ -0.0024 \pm 0.0052 \pm 0.0022 $ \\
& 2005 & CDF~\cite{CDF2005}     & $ +0.010  \pm 0.013  \pm 0.006  $ \\
& 2002 & CLEO~\cite{ycp_cleo}   & $ +0.019  \pm 0.032  \pm 0.008  $ \\
& 2000 & FOCUS~\cite{FOCUS2000} & $ +0.048  \pm 0.039  \pm 0.025  $ \\
& 1998 & E791~\cite{E7911998}   & $ -0.049  \pm 0.078  \pm 0.030  $ \\
\hline
{\boldmath $D^0 \to \pi^0\pi^0$} &
  2001 & CLEO~\cite{CLEO2001}  & $ +0.001  \pm 0.048 $ (stat. and syst. combined) \\
\hline
{\boldmath $D^0 \to K_s^0\pi^0$} &
  2001 & CLEO~\cite{CLEO2001}  & $ +0.001 \pm  0.013 $ (stat. and syst. combined) \\
\hline
{\boldmath $D^0 \to K^+K^-$} &
 2008 & Belle~\cite{Belle2008_2} & $ -0.0043 \pm 0.0030 \pm 0.0011 $ \\
& 2008 & BABAR~\cite{BaBar2008} & $ +0.0000 \pm 0.0034 \pm 0.0013 $ \\
& 2005 & CDF~\cite{CDF2005}     & $ +0.020  \pm 0.012  \pm 0.006  $ \\
& 2002 & CLEO~\cite{ycp_cleo}   & $ +0.000  \pm 0.022  \pm 0.008  $ \\
& 2000 & FOCUS~\cite{FOCUS2000} & $ -0.001  \pm 0.022  \pm 0.015  $ \\
& 1998 & E791~\cite{E7911998}   & $ -0.010  \pm 0.049  \pm 0.012  $ \\
& 1995 & CLEO~\cite{CLEO1995}   & $ +0.080  \pm 0.061             $ (stat.) \\
& 1994 & E687~\cite{E6871994}   & $ +0.024  \pm 0.084             $ (stat.) \\
\hline
{\boldmath $D^0 \to K^0_sK^0_s$} &
 2001 & CLEO~\cite{CLEO2001}   & $ -0.23  \pm 0.19  $ (stat. and syst. combined) \\
\hline
{\boldmath $D^0 \to \pi^+\pi^-\pi^0$} &
   2008 & BABAR~\cite{BaBar2008b} & $ -0.0031 \pm  0.0041 \pm  0.0017$ \\
&  2008 & Belle~\cite{Belle2008}  & $ +0.0043 \pm  0.0130 $ \\
&  2005 & CLEO~\cite{CLEO2005}  & $ +0.001^{+0.09}_{-0.07} \pm  0.05 $ \\
\hline
{\boldmath $D^0 \to K^+ K^-\pi^0$} &
   2008 & BABAR~\cite{BaBar2008b} & $ 0.0100 \pm  0.0167 \pm  0.0025$ \\
\hline
{\boldmath $D^0 \to K^-\pi^+\pi^0$} &
  2007 & CLEOc~\cite{CLEOc2007} & $  +0.002  \pm 0.004  \pm 0.008 $ \\
&  2001 & CLEO~\cite{CLEO2001a}  & $ -0.031   \pm 0.086 $ (stat.) \\
\hline   
{\boldmath $D^0 \to K^+\pi^-\pi^0$} &
  2005 & BELLE~\cite{Belle2005} & $ -0.006  \pm 0.053  $ (stat.) \\
&  2001 & CLEO~\cite{CLEO2001b}  & $ +0.09^{+0.25}_{-0.22}  $ (stat.) \\
\hline
{\boldmath $D^0 \to K^0_s\pi^+\pi^-$} &
 2004 & CLEO~\cite{CLEO2004}    & $ -0.009  \pm 0.021^{+0.016}_{-0.057} $ \\
\hline
{\boldmath $D^0 \to K^+\pi^-\pi^+\pi^-$} &
  2005 & BELLE~\cite{Belle2005} & $ -0.018  \pm 0.044  $ (stat.) \\
\hline
{\boldmath $D^0 \to K^+K^-\pi^+\pi^-$} &
  2005 & FOCUS~\cite{FOCUS2005} & $ -0.082  \pm 0.056  \pm .047  $ \\
\hline                   
\end{tabular}}
\end{center}
\caption{\cp\ asymmetry 
$A^{}_{CP}=[\Gamma(D^0)-\Gamma(\dbar)]/[\Gamma(D^0)+\Gamma(\dbar)]$
for $D^0,\dbar$ decays.}
\label{tab:cp_neutral}
\end{table*}


\subsubsection{Spectroscopy via Dalitz-plot analyses}

   A very high statistics charm experiment can hope to unravel many
   mysteries related to resonances in the 1-2~GeV region. This is
   because charm hadron masses are in the 2~GeV range and $\pi\pi$, $K\pi$ and
   $KK$ resonances often dominate charm particle decays. 
   Below we describe two categories of measurements that could 
   be done at a Tevatron charm experiment.
   In addition, there is much to be learned from decays of the
   $J/\psi, \psi(2S)$, and even the $\eta_c$, all of which 
   would be copiously produced in a Tevatron charm experiment.


\paragraph{Improvement in parameterizations of resonances} 

   Thus far, experiments have used mainly a Breit-Wigner functional form to 
   describe resonances, with some modifications for barrier penetration factors,
   etc. However, there is no well-established theory that prescribes a
   precise form for the propagator of wide resonances. Hence several 
   deviations from simple forms have been proposed.
   For example, Gounaris and Sakurai~\cite{GounarisSakurai68} have 
   proposed a formula for the case of the wide $\rho(770)$ resonance.
   A well-known success is the Flatte formula~\cite{Flatte76} for a
   coupled-channel description of the $f^{}_0(980)$. 
   With regards to describing scalar resonances, both $K$-matrix and 
   $P$-vector formalisms~\cite{Wigner46,WignerEisenbud47,Aitchison72} have 
   been proposed. 
   The $K$-matrix method is attractive because it preserves unitarity,
   but the goodness-of-fit obtained is often no better than that obtained
   using a simple sum of resonances, and the $K$ matrix itself contains 
   implicit assumptions.
   Another issue is whether Zemach formalism~\cite{Zemach64} or
   helicity formalism~\cite{Kopp01,Lau07} correctly describes decays.



   A scan of the particle data table of light, unflavored mesons 
   shows that, beyond a mass of around 1\gevm, one or more of the 
   mass, width, and major branching fractions of most resonances 
   are not well-known. 
   The parameters of the $f^{}_0(980)$ are not well-established. 
   This is also true for strange mesons apart from the $K^*(892)$: 
   the $K_2^*(1430)$, $K_3^*(1780)$, and $K_4^*(2045)$. 
   Other poorly measured or otherwise controversial 
   states~\cite{PDG08_scalars,PDG08_eta}
   include the $\sigma(600)$, $\kappa(800)$, $a^{}_0(980)$,
   $\eta(1295)$, $\eta(1440)$, $f^{}_1(1420)$, and $f^{}_1(1510)$.
   A charm experiment at the Tevatron could clarify
   whether all these states exist and, if so, measure 
   their parameters with much improved precision.

\paragraph{Spectroscopy via production ({\it e.g.}, double charm baryons)}
  
Doubly-charmed baryons were discovered~\cite{selex02} at Fermilab in 
forward hadroproduction with baryon beams.  Several states have been 
reported, each in several decay modes.  However, there has not yet 
been an independent confirmation of these states. A future Tevatron 
charm experiment would be able to confirm and study these states with 
a much larger data set than that used previously.



\subsection{Overview of new technologies}
  
\subsubsection{Silicon pixel detectors/vertexing}

Silicon pixel detectors will play a crucial role in a new high-rate 
fixed-target charm experiment.  Their contributions include pattern 
recognition in complex event topologies, radiation-hard high-rate 
capability so that the primary beam can go through the detector without 
compromising performance, and excellent spatial resolution enabling 
the reconstruction of interaction and decay points from measured 
charged particle tracks.

Historically, silicon microstrip detectors have played an important role in 
fixed-target charm experiments. When these high precision vertex detectors 
were introduced in the eighties, they revolutionized the study of heavy 
flavors. Besides offering high precision tracking and vertex information, 
they lead to the possibility of high statistics experiments, something that 
earlier generations of experiments, using bubble chambers or emulsions could 
not possibly accomplish. In 1985/1986, CCDs were used in a fixed-target 
charm experiment, the first application of pixel devices in high energy 
physics. Since then, silicon strip detectors have become major tracking 
elements in all collider experiments: for the Tevatron, LEP, B-factories, 
and now the LHC. CCDs were limited to only $e^+e^-$ colliders because of their 
readout speed.  On the other hand, hybridized pixel detectors, 
in which 
readout chips were bump-bonded to silicon 
sensors, have been used in heavy ion experiments at CERN (WA97, NA62) and 
are now being employed as the vertex detector for ATLAS, CMS, and ALICE. 
With the development and experience gained over the last decade or so, 
the hybridized pixel detector technology has matured, and certainly can 
be an important tool for future fixed-target charm experiments.

Pixel detectors offer excellent 
three-dimensional information, which leads to much 
better pattern recognition, avoiding ambiguities and ghost tracks. 
Its advantages over the 
two-dimensional information provided by the 
silicon strip detectors have been demonstrated by both the fixed-target 
experiments at CERN and also at SLD. With a pixel size of 
50 microns by 400 microns, test beam results achieved a resolution 
of better than 2 microns. The detector noise is about 100 electrons 
or less. This means such a detector would give a signal-to-noise ratio of better 
than 200:1.  These detectors are also very quiet, and the spurious hits, 
as observed during the commissioning phase of the LHC experiments, are 
of order of $10^{-5}$. Furthermore, such devices can be self-triggered. 
All the readout chips used in the LHC experiments have the feature 
of being data-driven, which means that the chip generates a fast 
signal 
when a hit is registered above threshold. ALICE 
has used this information, and has taken a lot of cosmic ray and 
first beam data using a pixel-detector trigger.  

Pixel detectors, because of their fine segmentation, can also handle 
very high rate, and handle high radiation dosage.  These devices have all the 
excellent features that are required in a next generation of 
charm experiments.

Since 1998, Fermilab has been active in the pixel R\&D effort.  
This has led to the development of the FPIX series of pixel 
readout chips for the BTeV experiment. When BTeV was cancelled, 
a group from Los Alamos picked up the design and used the chip, 
sensor, interconnect, and a lot of the mechanical design to build 
two forward muon stations for the PHENIX experiment.  
With small modifications, such a design could be well suited 
for a new charm experiment at the Tevatron.
 
\subsubsection{Triggering on decay vertices, impact parameters}

With the technical advances in detectors and electronics made
since the last Fermilab fixed-target experiments, it is now possible 
to build a high-rate trigger system that selects charm events at the
lowest trigger level by taking 
advantage of the key property that differentiates charm particles from 
other types of particles, namely their characteristic lifetimes. To achieve
this, a new experiment would trigger on charm decay vertices by performing track 
and vertex reconstruction to search for evidence of a particle-decay vertex 
that is located tens to thousands of microns away from a primary 
interaction vertex. In practice, this would be done by reconstructing
primary vertices and selecting events that have additional tracks with 
large impact parameters with respect to the primary vertex. The main 
advantage of this approach is that it suppresses light-quark background 
events while retaining high efficiency for charm events at the first 
stage of triggering by maximizing the trigger acceptance compared to 
trigger strategies that rely on detecting specific final-state particles, 
such as muons, or selecting events based on $E^{}_T$ cuts.

A trigger and data acquisition system for a new charm experiment would be 
able to take advantage of what has been learned from other experiments. While 
the power of silicon strip detectors for tracking and vertex reconstruction has 
been demonstrated by numerous experiments, it is the high-resolution 
three-dimensional tracking capability provided by a pixel vertex detector 
that permits a straightforward design for triggering on detached vertices
at the first stage of a trigger system. A pixel vertex detector together 
with zero-suppressed readout of the data provide what is needed to perform 
the pattern recognition, track reconstruction, vertex reconstruction,
and impact-parameter calculations that form the basis for a detached-vertex 
trigger. The design of the BTeV experiment's trigger was based on the following features:
\begin{itemize}
\item field programmable gate arrays (FPGAs) or comparable devices 
for pattern recognition;
\item low-cost memory to buffer event data and allow for relatively 
long latencies in the first-level trigger;
\item commodity off-the-shelf (COTS) networking and processing hardware.
\end{itemize}

BTeV demonstrated the possible tradeoffs between calculations performed 
by FPGAs and general-purpose processors. In the BTeV trigger FPGAs performed 
most of the pattern recognition for pixel data, since FPGAs excel at 
performing large numbers of rudimentary calculations in parallel.
The remaining calculations were performed by general-purpose processors. 
One of the key features of the BTeV trigger was flexibility in the 
design that made it possible to move calculations performed in
processors into FPGA hardware, thereby improving performance and 
reducing the cost of trigger hardware. Several FPGA-based algorithms 
were developed at Fermilab that could also be applied to a new 
charm experiment. Examples include an FPGA-based track segment 
finder and a fast ``hash sorter'' that sorted track-segment data 
before sending it to a general-purpose processor.

BTeV also demonstrated that advances in electronics make it possible to 
build a data acquisition system that will buffer event data long enough for 
a first-level trigger to analyze every interaction and perform complex 
operations to search for evidence of a detached vertex. The BTeV trigger 
design included enough memory to buffer data from the entire detector for 
approximately 800 ms, which was over three orders of magnitude more than 
the average processing time required by the first-level trigger. 
In addition to the large event buffer, the BTeV design relied on 
commodity networking and processing hardware to implement a 
sophisticated detached-vertex trigger that could be built for 
a reasonable cost. The key features of this design are being 
considered by the LHCb Collaboration for their upgrade in
the middle of the next decade.

The CDF experiment has been using a decay-vertex trigger at the
second level~\cite{CDF_SVT} to record large samples of two-body
$B$ and $D$ decays. This success demonstrates the feasibility and
capability of a heavy-flavor-decay trigger used in a hadroproduction 
environment.

\subsubsection{RICH detectors, $\pi/K$ separation} 

The physics goals of a fixed-target charm experiment require good charged 
particle identification to observe various decay modes of interest. At the 
Tevatron fixed-target energies, one must be able to separate pions, kaons, 
and protons with high efficiency over a range of momentum from several 
GeV up to hundreds of GeV. This can be accomplished by using a Ring Imaging 
Cherenkov detector (RICH).

From the early days of the OMEGA experiment, over the years, RICH detectors 
have been built and operated in different environments. They were used in 
fixed-target experiments at Fermilab ({\it e.g.} E665, E706, E789, E781), 
in \hb\ at DESY, as well as in $e^+e^-$ collider experiments (CLEO, DELPHI and SLD).
Currently, a RICH detector is used in a hadron collider experiment (LHCb). 

The detector performance and cost is determined, to a large extent, by the 
choice of the photo-detector. 
In the early days, experiments used gas detectors based on photo-ionizing 
gas such as TMAE or TEA. 
Operationally, this has not been easy. On the other hand, the new rounds 
of experiments tend to use 
commercial detectors such as PMT (SELEX), MAPMT (HERA-B) and HPD (LHCb) 
which offer stability, ease of operation, and maintenance at a moderate cost.

We can take the SELEX RICH as an example. The RICH vessel is 10.22~m long, 
93 inches in diameter and filled with neon at atmospheric pressure. At the 
end of the vessel, an array of 16 hexagonal mirrors are mounted on a low-mass 
panel to form a sphere of 19.8~m in radius. Each mirror is 10~mm thick, 
made out of low-expansion glass. For the photo-detector, SELEX used 
2848 0.5-inch photomultiplier tubes arranged in an array of 89~$\times$~32. 
Over a running period of 15 months, detector operation was very stable. 
The ring radius resolution was measured to be 1.56~mm and, on average, 
13.6 photons were observed for a $\beta=1$ particle.  

\subsubsection{Micropattern gaseous tracking detectors, {\it e.g.\/} MSGC, GEM, and Micromega}

Previous generations of fixed-target charm experiments typically used 
large area gaseous detectors ({\it e.g.} drift chambers or multiwire gaseous 
chambers) for charged-particle tracking purposes.  In high rate environments, 
these detectors suffered from inefficiency. In extreme environments, like 
regions around the incident beam, there was a dead region. For example, 
in the charm E-791 experiment at Fermilab, there was a large drift-chamber 
inefficiency (``hole'') around the beam line which had to be constantly 
monitored and corrected in the Monte Carlo acceptance calculations. 
This also led to significant loss in the overall efficiency of the 
spectrometer.

Since the early 1990's, there have been substantial advances in 
micropattern gaseous detectors. These include MSGC (multistrip 
gaseous chamber), GEM (gaseous electron multiplier), and Micromega 
devices. Currently, the state-of-the-art is that chambers as large 
as $40 \times 40$~cm$^2$ can be built using either GEMs or Micromegas. 
These type of detectors have been operated reliably in the last 
generation of high rate fixed-target experiments such as COMPASS.

In a future high-rate heavy-flavor experiment, one can build a set 
of Micromegas or GEM detectors near the beam region to handle the 
high rate. Outside this region, the more conventional drift chambers 
can be used. This will allow operation at high rates with large area coverage.

\subsection{Summary}

In summary, we note the following and conclude:
\begin{itemize}
\item $D^0$-$\dbar$ mixing is now established, and attention
has turned to the question of whether there is \cpv\ in this system. 
\item Technical advances in detectors and electronics made
since the last Fermilab fixed-target experiments ran would make
a new experiment much more sensitive to mixing and \cpv\ effects.
Silicon strips and pixels for vertexing are well-developed, and 
detached-vertex-based trigger concepts and prototypes exist 
({\it e.g.} \hb, CDF, BTeV, LHCb).
\item Such an experiment would have substantially better sensitivity
to mixing and \cpv\ than all \belle\ and \babar\ data together 
will provide. The Tevatron data should have less background than 
LHCb data. Systematic uncertainties may also be less than those 
of 
any Super-$B$ Factory 
experiments and LHCb.
\item The Tevatron and requisite beamlines are essentially available.
\item Such an experiment could help untangle whatever signals for new 
physics appear at the Tevatron or LHC.
\end{itemize}

Recently, a working group has formed to study the physics potential of a charm experiment at the Tevatron in more detail. Information about this working group and its results can be obtained at \textit{http://www.nevis.columbia.edu/twiki/bin/view/FutureTev/WebHome}.

In brief, we write this chapter to keep the possibility of a 
fixed-target charm experiment at the Tevatron a viable option 
for Fermilab (and the broader international HEP program), to be decided upon once 
there is a clearer picture of available funding, manpower, and 
feasibility of the current roadmap.

\section{Neutrino-electron Scattering}

Neutrino-electron scattering ($\nu_\mu + e \rightarrow \nu_\mu + e$)
is an ideal process to search for beyond the Standard Model physics at
Terascale energies through precision electroweak measurements.  The
low cross section for this process demands a very high intensity beam.
In order to reduce systematics and reach a cross-section precision better than 1\%, this process can be normalized to its
charged-current sister, ``inverse muon decay'' ($\nu_\mu + e
\rightarrow \nu_e + \mu$).  The threshold for this interaction is 11~GeV.
Therefore, the experiment requires a high energy neutrino flux, as can only
be provided by a $\sim$1~TeV primary proton beam.  Once a
high-energy, high-intensity neutrino flux is established, a detector
optimized for $\nu-e$ scattering can also be used for precision
structure function and QCD measurements and direct searches.

The physics reach of NuSOnG for beyond the Standard Model physics is 
in the 1~to~7~TeV range, depending on the model.   The sensitivity to new
physics complements the LHC and brings unique new opportunities to the
program.   The full physics program is discussed in detail elsewhere~\cite{NuSOnGEW,NuSOnGQCD,EOI}. In this paper, we present an experimental overview 
which illustrates the value of this endeavor.
\subsection{The beam}

\begin{figure}[tb]
\begin{centering}
\begin{tabular}{cc}
\includegraphics[height=2.6in]{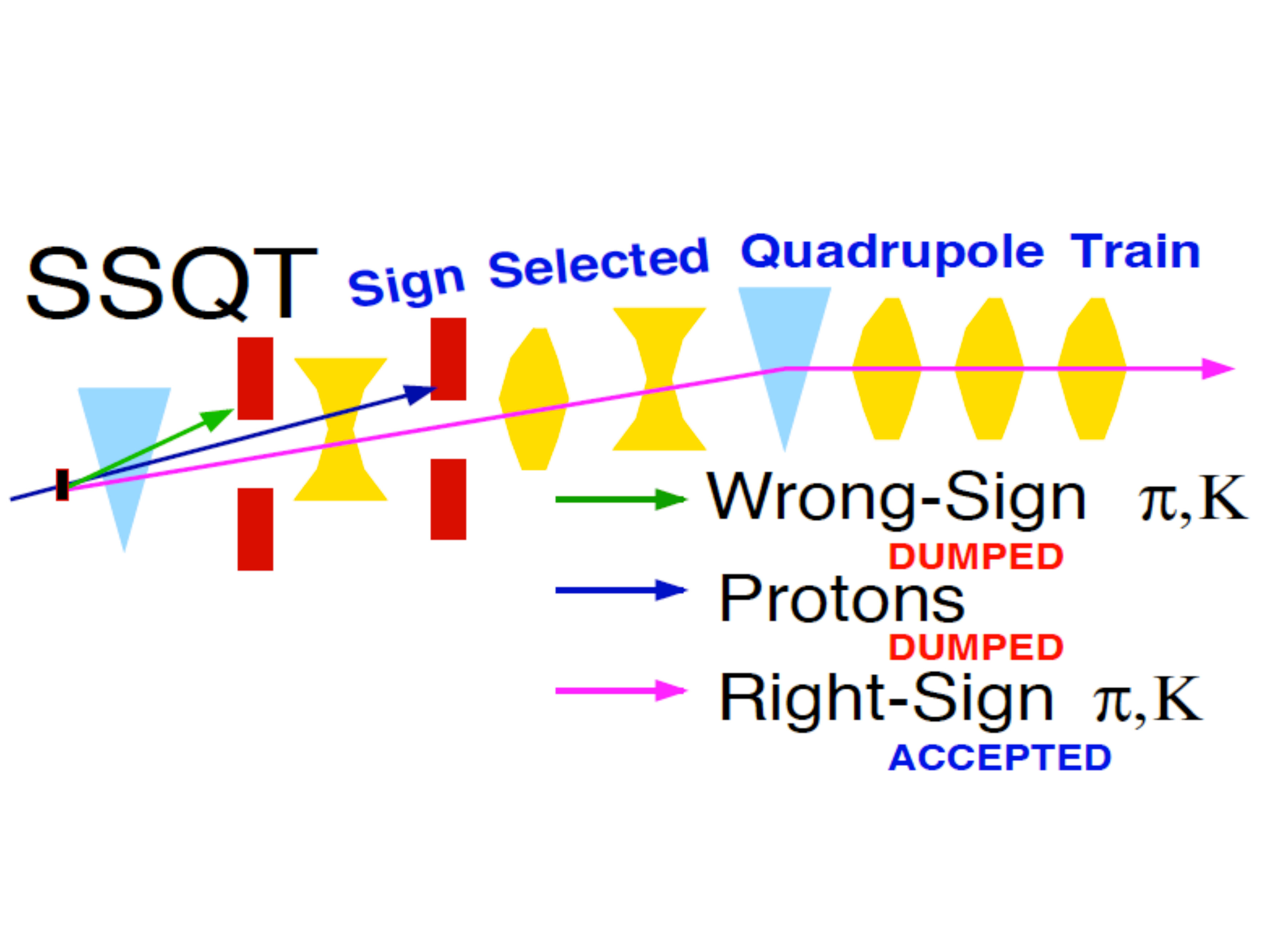} &
\end{tabular}
\vspace{-1.cm}
\caption{The NuSOnG beam design, identical to that used by the NuTeV experiment. From Ref.~\protect\cite{NuTeVbeam}.}
\label{ssqt}
\end{centering}
\end{figure}

For this discussion, we will assume a NuSOnG beam design which is the
same as that used by the NuTeV experiment (see Fig.~\ref{ssqt}), which ran from 1993-1996 at
Fermilab~\cite{NuTeVbeam}. We will assume $2\times 10^{20}$ high
energy (800~GeV to 1~TeV) protons impinge on a beryllium oxide target.
The resulting mesons traverse a quadrupole-focused, sign-selected
magnetic beamline, hence the design is called a ``sign-selected quad
triplet'' or SSQT.  NuSOnG will run with 1.5$\times 10^{20}$ protons on target in neutrino mode, and $0.5 \times 10^{20}$ protons on target in antineutrino
mode.  The result is a beam of very ``right sign'' purity ($>98\%$)
and low $\nu_e$ contamination ($2\%$).  The $\nu_e$ in
the beam are due mainly to $K^+$ decays which can be well-constrained
by the $K^+ \rightarrow \nu_\mu$ flux which populates the high energy
range of the neutrino flux.  The magnetic bend substantially reduces
$\nu_e$ from $K_L$ decay which tend to go forward and will thus not be
directed at the detector.
\subsection{The detector: segmented glass and LAr options}

The baseline design for the NuSOnG detector is a 3.5~kton glass-target
design inspired by the design of the Charm II experiment.  The
detector is broken into four identical subdetectors, each consisting
of a $5{\rm~m}\times 5{\rm~m}\times 29 {\rm~m}$ target plus toroid muon spectrometer.  Breaking the design into four
sections assures high acceptance for muons produced in the target
calorimeter to reach the toroid.  A gap of 15~m extends between
each detector to allow for a calibration beam to impinge on the target.
The total length of the detector is, therefore, 200 m.

The total target is composed of 2500 sheets of glass which are 2.5 cm
(0.25 $\lambda_0$) thick.  This provides an isoscalar target for
neutrino-quark interaction studies.  Interspersed between the glass
sheets are proportional tubes or scintillator planes.  
The total target mass is six times greater than NuTeV.

The signal processes are: $\nu_\mu + e \rightarrow \nu_\mu +e$ and
$\nu_\mu + e \rightarrow \mu + \nu_e$.  These must be distinguished
from the background processes of $\nu_e + n \rightarrow e + p$ and
$\nu_\mu + n \rightarrow \mu + p$.  These processes become
background for certain kinematic cases when the proton is not detected.  
In the initial studies for NuSOnG, which was
designed as interleaved one-radiation length glass targets and live
detectors, a large systematic error came from the number of background
events where the proton was lost in the glass~\cite{NuSOnGEW}.
Protons may be lost in the glass because they are produced at
relatively wide angles and low energies.    Motivated by 
this, the collaboration has been considering other designs
for a target calorimeter.

\begin{figure}[tb]
\begin{centering}
\begin{tabular}{cc}
\hspace{-1.5cm}
\includegraphics[height=2.6in]{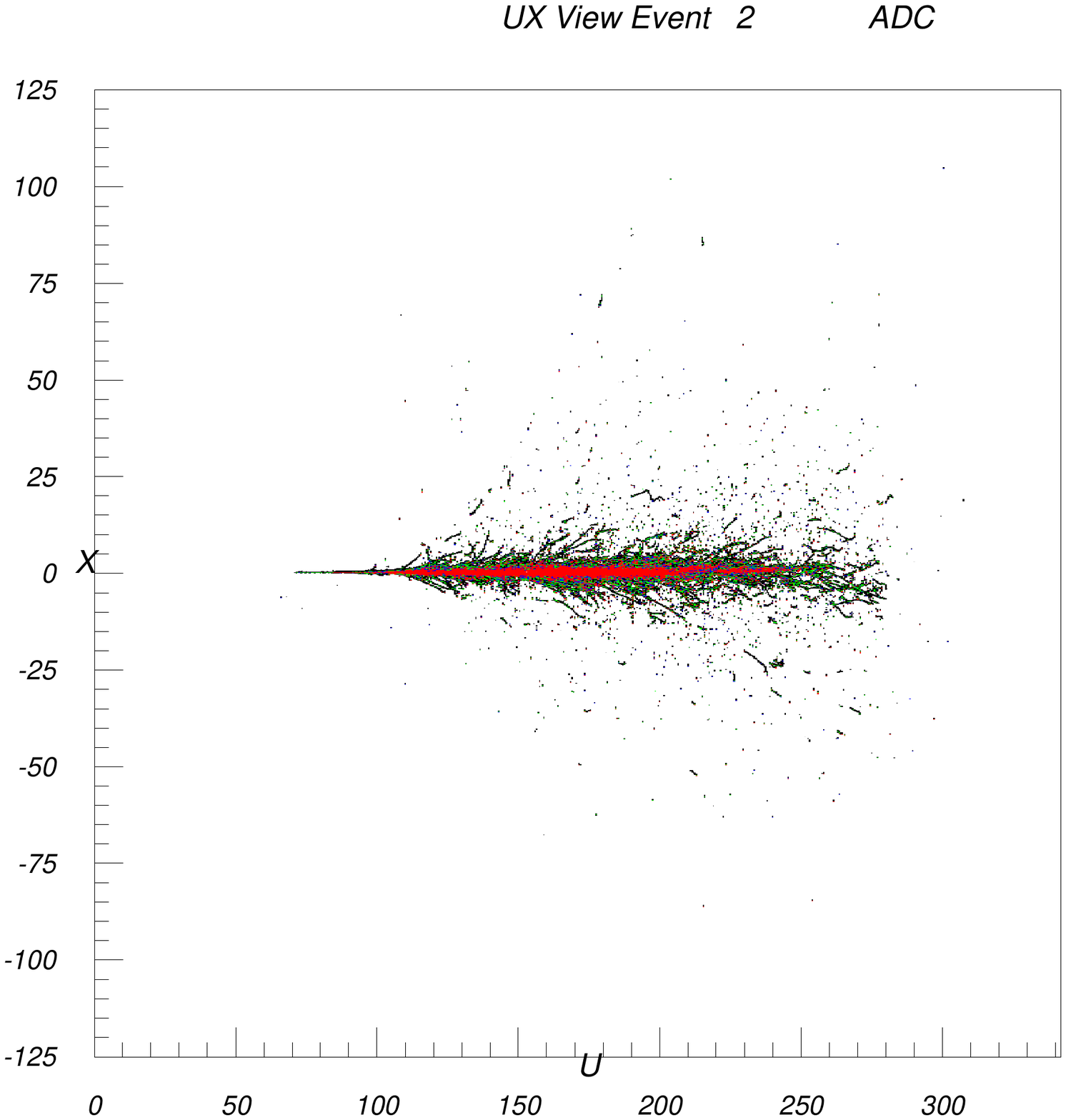} &
\hspace{-3.cm}
\includegraphics[height=2.6in]{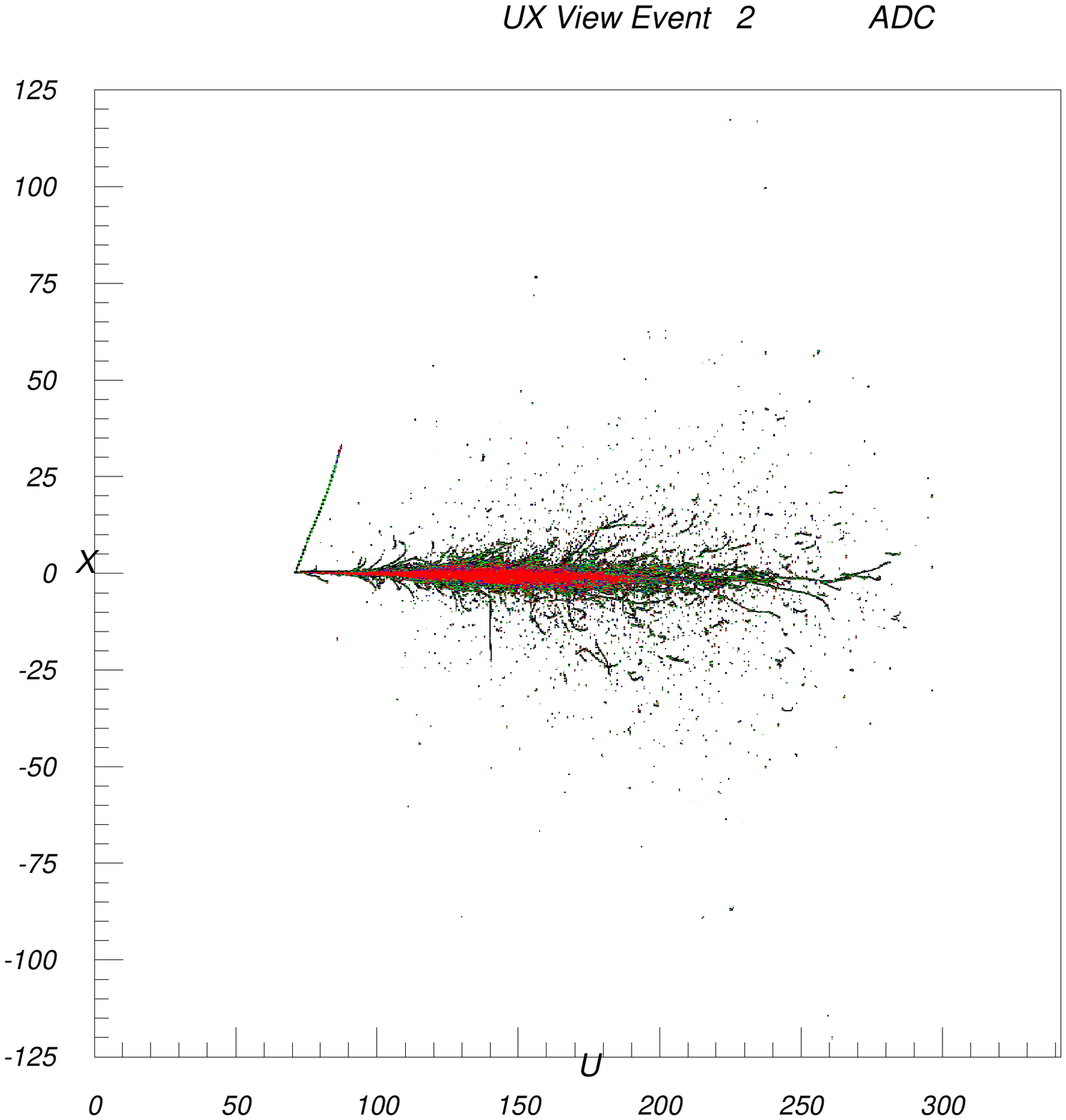} 
\end{tabular}
\vspace{-.5cm}
\caption{A 60~GeV $\nu-e$ neutral current scatter in an LAr detector (left) and a typical 60~GeV $\nu_e + n$ background event (right).}
\label{nue}
\end{centering}
\end{figure}

LArTPC detectors provide a fully-live alternative in which the proton
signature in background events is easily identified.  Fig.~\ref{nue}
compares a 60~GeV $\nu-e$ neutral current scatter in an LAr detector to a typical 60~GeV $\nu_e + n$ background event.  One
can see that a proton track, which is at a large angle relative to the
shower, is clearly visible.  A 2~kton LArTPC detector is expected to
have similar sensitivity to the 3.5~kton glass NuSOnG detector.
An LArTPC would require substantially less electronics than the glass
detector, and should be proportionately less expensive.  

The NuSOnG LArTPC alternative is very similar in design to the
technology for the $\nu_\tau$ physics discussed later in this paper. The similarity between an LAr-based
NuSOnG and a future $\nu_\tau$ detector is demonstrative of the synergy within this 
overall fixed-target program.

\subsection{Neutral \& charged current processes}

Remarkable rates are acquired when the 3.5~kton detector is combined with the high intensity, high
energy beam.  One expects $>600$M
$\nu_\mu$ CC events and $>65$M $\nu_e$ CC events.  This can be
compared to past samples of $<20$M~\cite{CCFR,NuTeV,CDHS,CHARMII,NOMAD,CHORUS,SuperK,K2K,MINOSnumu} and
$\sim 500$k~\cite{CCFR,NuTeV,CDHS,CHARMII,NOMAD,CHORUS,SuperK,K2K,MINOSnumu,SNO,Bugey,Chooz,MUNU},
respectively.  With such large data samples, NuSOnG can search for
processes which are within the Standard Model, but rare, or beyond the
Standard Model which have not been studied before.

The expected rates for specific event types are given in 
Table~\ref{Tab:rates}.   In particular, one should note that the $\nu-e$ scattering sample is 40 times larger than that of previous experiments.

\begin{table}[h]
\centering
{\begin{tabular}{c|c}
600\/M & $\nu_\mu$ CC Deep Inelastic Scattering\\
190\/M & $\nu_\mu$ NC Deep Inelastic Scattering \\
75\/k & $\nu_\mu$ electron NC elastic scatters (ES) \\
700\/k &$\nu_\mu$ electron CC quasi-elastic scatters (IMD) \\
33\/M &  $\bar \nu_\mu$ CC Deep Inelastic Scattering \\
12\/M & $\bar \nu_\mu$ NC Deep Inelastic Scattering \\
7\/k & $\bar \nu_\mu$ electron NC elastic~scatters (ES)\\
0\/k &  $\bar \nu_\mu$ electron CC quasi-elastic scatters (WSIMD)\\
\end{tabular}}
\caption{Number of events in NuSOnG assuming $2\times 10^{20}$ protons on target. NC indicates ``neutral current''and CC indicates ``charged current.''}
\label{Tab:rates}
\end{table}

The high event rates for neutrino neutral current scattering provide
a remarkable opportunity to probe for new physics through the weak
mixing angle, $\sin^2 \theta_W$, and the ratio of neutral to charged
current couplings, $\rho$.  This physics can be accessed through four
modes: $\nu_{\mu}+e^- \rightarrow \nu_{\mu}+ e^-$, $\bar \nu_{\mu}+e^-
\rightarrow \bar \nu_{\mu}+ e^-$, $\nu_{\mu}+q \rightarrow \nu_{\mu}+
q$, and $\bar \nu_{\mu}+q$.  There has been a long history of
experiments which have exploited precision neutral current quark
scattering, but the ultra-high rates for neutrino-electron scattering
are a new opportunity raised by the high-energy, high-intensity primary
beam.  A deviation from the Standard Model predictions in both the
electron and quark measurements would present a compelling case for
new physics.

An essential feature to the NuSOnG $\nu-e$ study is that the NC event
rate can be normalized to the CC process, called ``inverse muon
decay'' (IMD), $\nu_{\mu}+ e^- \rightarrow \nu_e + \mu^-$.  This
process is well understood in the Standard Model due to precision
measurement of muon decay~\cite{mudk}.  Since the data samples are
collected with the same beam, target, and detector at the same time,
the ratio of ES to IMD events cancels many systematic errors while
maintaining a strong sensitivity to the physics of interest.  Our
measurement goal of the ES to IMD ratio is a 0.7\% error, adding
systematic and statistical errors in quadrature~\cite{NuSOnGEW}.  The
high sensitivity which we propose arises from the combined high energy
and high intensity of the NuSOnG design, leading to event samples more
than an order of magnitude larger than past experiments.

Normalizing the ES to the IMD events -- which can only occur because
of the TeV-scale primary beam -- represents a crucial step forward
from past ES measurements, which have normalized neutrino-mode ES
measurements to antineutrino mode, $\bar \nu_{\mu}+e^- \rightarrow
\bar \nu_{\mu}+ e^-$~\cite{CHARMII,ahrens}.  In fact, the level
of precision expected from NuSOnG cannot be reached in lower energy
experiments using antineutrino normalization.  The improvement from
the NuSOnG method is in both the experimental and the theoretical
aspects of the measurement.  First, the flux contributing to IMD and
$\nu$ES is identical, whereas neutrino and antineutrino fluxes are
never identical and so require corrections.  Second, the ratio of
$\nu$ES to $\bar \nu$ES cancels sensitivity to beyond Standard Model
physics effects from the NC to CC coupling ratio, $\rho$, which are
among the primary physics goals of the NuSOnG measurement.  In
contrast, there is no such cancellation in the ES to IMD ratio.

\subsection{Beyond Standard Model reach}

Elastic neutrino electron scattering is a purely leptonic electroweak
process. It can be computed within the Standard Model with high
precision~\cite{'tHooft:1971ht} and hence provides a very
clean probe of physics beyond the Standard Model. The effect of new,
heavy ($M_{\rm new}\gg \sqrt{s}$) degrees of freedom to
$\nu_{\mu}e^-\to\nu_{\alpha}e^-$, where $\alpha=e,\mu,\tau$ can be
parameterized by the effective Lagrangian
\begin{equation}
\mathcal{L}_\mathrm{NSI}^{e} = +\frac{\sqrt{2}}{\Lambda^2}
\Bigl[\, \bar{\nu}_\alpha \gamma_\sigma P_L \nu_\mu 
\,\Bigr]
\Bigl[\,
 \cos\theta\, \bar{e}\gamma^\sigma P_L e
+\sin\theta\, \bar{e}\gamma^\sigma P_R e
\,\Bigr]  \;
\label{eq:leff}
\end{equation}
New physics, regardless of origin\footnote{We are neglecting
  neutrino currents involving right-handed neutrinos or lepton-number
  violation. These are expected to be severely suppressed as they are
  intimately connected to neutrino masses (and, to a lesser extent,
  charged-lepton masses). Once constraints related to neutrino masses
  are taken into account, these contributions are well outside the
  reach of TeV-sensitive new physics searches.}, manifests itself
through two coefficients: $\Lambda$ and $\theta$. $\Lambda$ is the
mass scale associated with the new physics, while $\theta\in[0,2\pi]$
governs whether the new physics interacts mostly with right-chiral or
left-chiral electrons, and also governs whether the new physics
contribution interferes constructively or destructively with the
Standard Model process ($Z$-boson $t$-channel exchange) in the case
$\alpha=\mu$.

\begin{figure}[h]
\centering
\scalebox{0.5}{\includegraphics[clip=true]{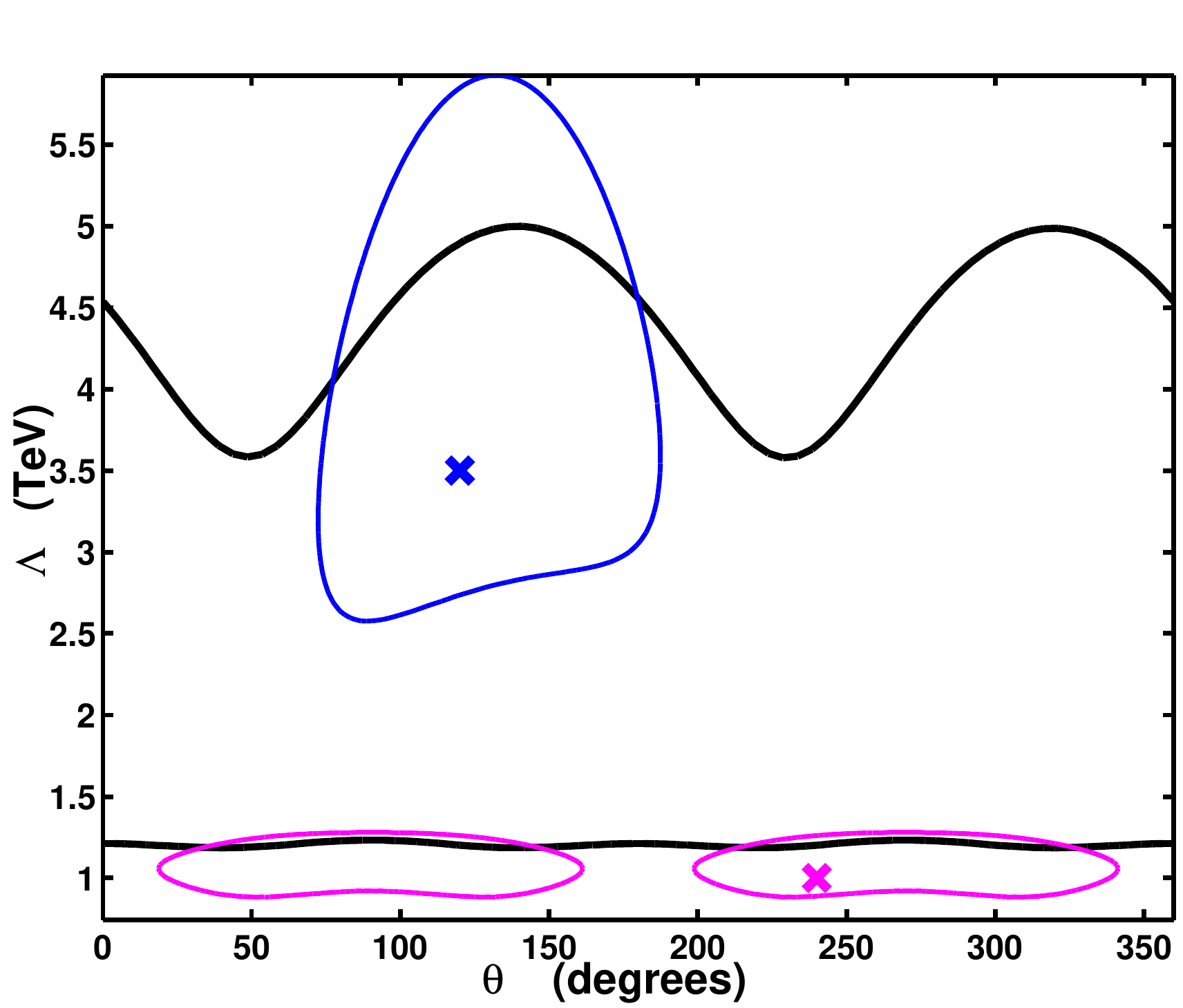}}
\vspace{1mm}
\caption{
Dark Lines: 95\% confidence level sensitivity of NuSOnG to new heavy  
physics described by Eq.~(\ref{eq:leff}) when $\nu_\alpha=\nu_{\mu}$  
(higher curve) and $\nu_\alpha\neq\nu_{\mu}$ (lower curve).  Closed  
Contours: NuSOnG measurement of $\Lambda$ and  $\theta$, at the 95\%  
level, assuming $\nu_\alpha=\nu_{\mu}$, $\Lambda=3.5$~TeV and $ 
\theta=2\pi/3$ (higher, solid contour) and $\nu_\alpha\neq\nu_{\mu}$, $ 
\Lambda=1$~TeV and $\theta=4\pi/3$ (lower, dashed contour). Note that  
in the pseudoelastic scattering case ($\nu_\alpha\neq\nu_{\mu}$) $, \theta$ and $\pi+\theta$ are physically indistinguishable. From Ref.~\protect\cite{NuSOnGEW}.}
\label{fig:eff_limit}
\end{figure}

Fig.~\ref{fig:eff_limit} depicts NuSOnG's ability to exclude $\Lambda$
as a function of $\theta$ for $\alpha=\mu$ or $\alpha\neq\mu$ assuming
its $\nu-e$ elastic scattering data sample is consistent with Standard
Model expectations. It also depicts NuSOnG's ability to measure
$\Lambda$ and $\theta$ in case a significant discrepancy is
observed. For more details see Ref.~\cite{NuSOnGEW}. In the case
$\alpha=\mu$, where new physics effects interfere with the Standard
Model contribution, NuSOnG is sensitive to $\Lambda\lesssim 4$~TeV
while in the $\alpha\neq \mu$ case NuSOnG is sensitive to
$\Lambda\lesssim 1.2$~TeV. The new physics reach of NuSOnG is
competitive and also complementary to that of the LHC, where new
physics in the neutrino sector is hard to access. The new physics
reach of NuSOnG is competitive with other leptonic probes (which involve
only charged leptons), including LEP2~\cite{lep2}, and precision
measurements of M{\o}ller scattering~\cite{moller}.

\begin{figure}[h]
\begin{center}
\scalebox{0.5}{\includegraphics{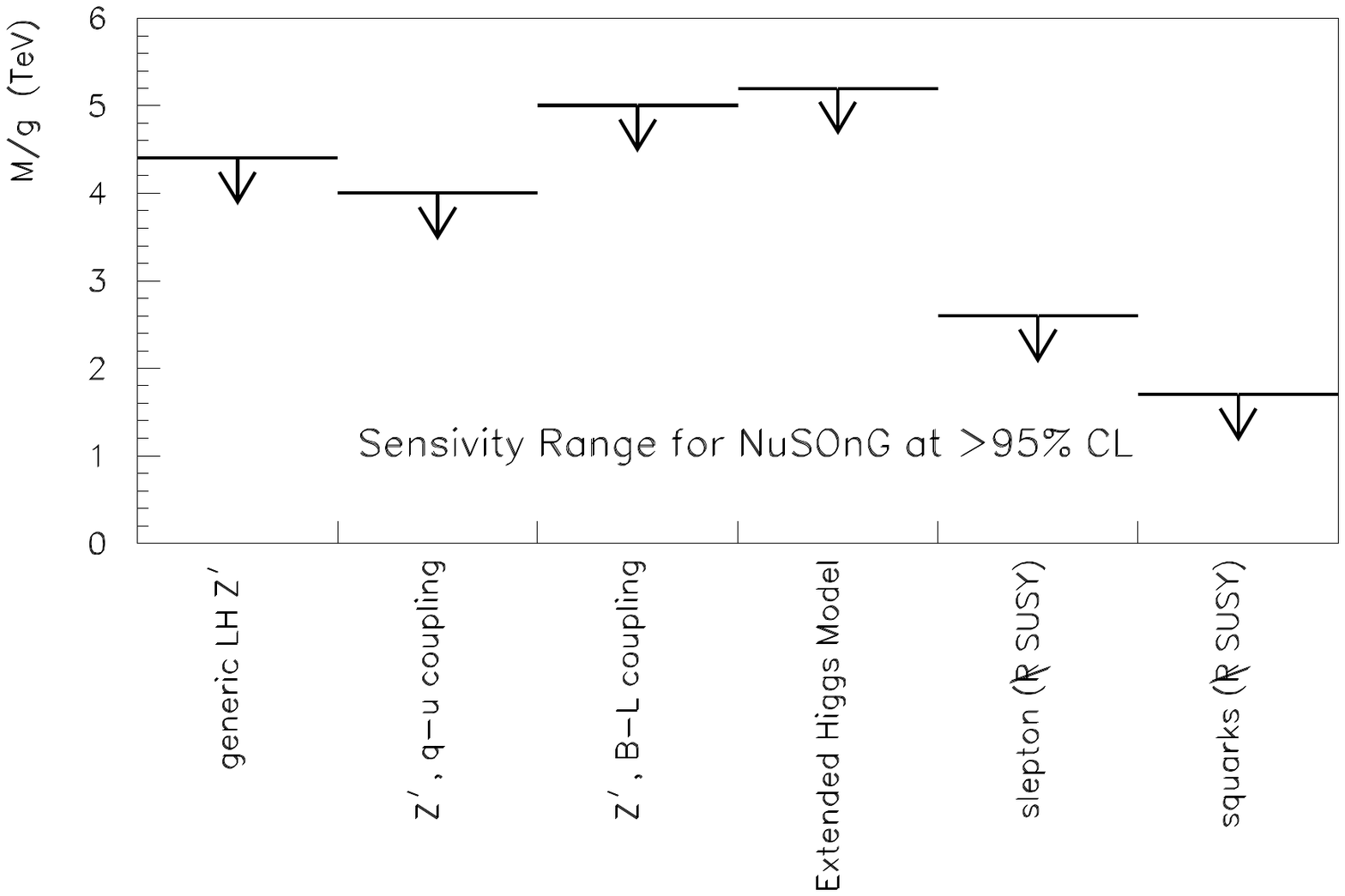}}
\vspace{-2.3in}
\caption{\label{massreach} Some examples of NuSOnG's 2$\sigma$
  sensitivity to new high-mass particles commonly considered in the
  literature.  For explanation of these ranges, and further examples,
  see Ref.~\protect\cite{NuSOnGEW}.}
\end{center}
\end{figure}

Several specific new physics scenarios can be probed by a high
statistics, high precision measurement of neutrino--matter
interactions. NuSOnG's reach to several heavy new physics scenarios is
summarized in Fig.~\ref{massreach}. There, we consider not only
information obtained from neutrino--electron elastic scattering and
inverse muon decay but also from neutrino--quark scattering (both
neutral current and charge current). If the new physics scale is
below a few TeV, we expect NuSOnG data to significantly
deviate from Standard Model expectations.

A more detailed comparison of NuSOnG's capabilities is summarized in Table~\ref{Tab:modeloverview}.
 
\begin{table*}[h] 
\centering
\footnotesize
{\begin{tabular}{|c|l|}
\hline
Model &   Contribution of NuSOnG Measurement \\ \hline \hline
Typical $Z^\prime$ Choices:  $(B-xL)$,$(q-xu)$,$(d+xu)$ &   At the level of, and complementary to, LEP II bounds.\\ \hline
Extended Higgs Sector & At the level of, and complementary to, $\tau$ decay bounds. \\ \hline
R-parity Violating SUSY & Sensitivity to masses $\sim2$~TeV at 95\% CL. \\
                        & Improves bounds on slepton couplings by $\sim30\%$ and \\
                        & on some squark couplings by factors of 3-5. \\ \hline
Intergenerational Leptoquarks (non-degenerate masses) & Accesses unique combinations of couplings.  \\ 
&  Also accesses coupling combinations explored by  \\
&  $\pi$ decay bounds, at a similar level.\\ \hline
\end{tabular}
\caption{Summary of NuSOnG's contribution in the case of specific models. See Ref.~\protect\cite{NuSOnGEW} for details.}
\label{Tab:modeloverview}}
\normalsize
\end{table*}

Finally, NuSOnG is also sensitive to the existence of new {\em light}
degrees of freedom, including neutral heavy leptons. A particularly
interesting signal to look for is wrong-sign inverse muon decay
($\bar{\nu}_{\mu}+e^-\to \bar{\nu}_{\alpha}+\mu^-$), which, given our
current understanding of neutrino masses and lepton mixing, only
occurs at a negligible level. Wrong-sign inverse muon decay would
point to short oscillation length neutrino oscillations mediated by
sterile neutrinos, a non-unitary lepton mixing matrix, or other
non-standard neutrino interactions.

In summary, NuSOnG is sensitive to a wide range of beyond Standard
Model physics and complementary to new physics which
might be observed at the LHC.  The program is also complementary to the
$\nu_\tau$ experiment and Neutral Heavy Lepton search described below.
This unique physics capability arises from the high-flux, high-energy
neutrino beam produced by primary protons at $\sim$1~TeV, which allows
normalization to IMD events for the first time.

Parallel to the studies summarized in this paper, an independent analysis of high-precision QCD topics possible with a new high-energy neutrino beam was carried out. These high-precision measurements are sensitive to Charge Symmetry Violations and other “new physics” processes that can significantly influence precision Standard Model parameter extraction.  In addition, the large statistics allows the separate extraction of n and n-bar structure functions leading to measurements of both DxF2 and DxF3 as well as RL for n and n-bar.  Finally, this high statistics QCD study will include large samples on different nuclear targets allowing us to disentangle the nuclear effects present in neutrino-nucleus processes that appear to be different than the nuclear effects in charged-lepton scattering. 

\section{$\nu_\tau$ Experiments}

Since the discovery of the charged $\tau$ lepton~\cite{Perl:1975bf}, physicists
have assumed the existence of a weak partner particle, $\nu_\tau$, analogous
to the neutrino partners of the $e$ and $\mu$ leptons as required by the
Standard Model and directly observed for the first time only recently by the
DONuT experiment~\cite{Kodama:2008zz,Kodama:2000mp}. Indeed, the wealth of
studies of charged $\tau$ lepton properties also require an accompanying
tau-neutrino ($\nu_\tau$) for a consistent description of the observed
dynamics. Little is directly known about the $\nu_\tau$
itself\footnote{We will henceforth mean by ``$\nu_\tau$'', the neutrino
initially prepared in the weak eigenstate with $\tau$ lepton number $\pm1$, as
the mass and weak eigenstates of the Standard Model neutrinos have been shown to be distinct with the observation of neutrino
oscillation.}. To date, only nine $\nu_\tau$ charged-current events have ever
been detected~\cite{Kodama:2008zz,Kodama:2000mp} and all other information we
have on this neutrino weak eigenstate is indirect\footnote{Solar and
atmospheric neutrino experiments have data which is best interpreted as
evidence for $\nu_{\tau}$ neutral current interactions. The Super-Kamiokande
atmospheric neutrino data also statistically favors the presence of both
neutral current and charged current $\nu_\tau$ initiated
events~\cite{Abe:2006fu,Fukuda:2000np}.}. An experiment sensitive to $\tau$
leptons placed along the path of an intense, $\nu_{\tau}$-rich neutrino beam
would add significantly to our understanding of electroweak interactions and
would be sensitive to certain hard-to-get manifestations of new physics.

Advances in the development of Liquid Argon Time Projection Chambers (LArTPCs),
notably for the ArgoNeuT~\cite{ArgoNeuThttp} and
MicroBooNE~\cite{MicroBooNEprop} Fermilab projects and
ICARUS-T600~\cite{Amerio:2004ze} at LNGS, suggest it would be a good choice of
base detector technology. A 1 kiloton LArTPC would fulfill the physics
requirements to discriminate high energy $\nu_\tau$ charged current
interactions while also providing a useful step in the development of LArTPC
technology. Experience with progressively larger LArTPC devices will enable
easier deployment for future projects with requirements for fiducial masses
of 5 kilotons or more, as has been suggested for future long-baseline
neutrino-oscillation experiments at the Deep Underground Science and
Engineering Laboratory (DUSEL)~\cite{DUSELhttp} and other facilities.

\subsection{The neutrino source}

\begin{figure}[tb]
\begin{centering}
\includegraphics[width=5in]{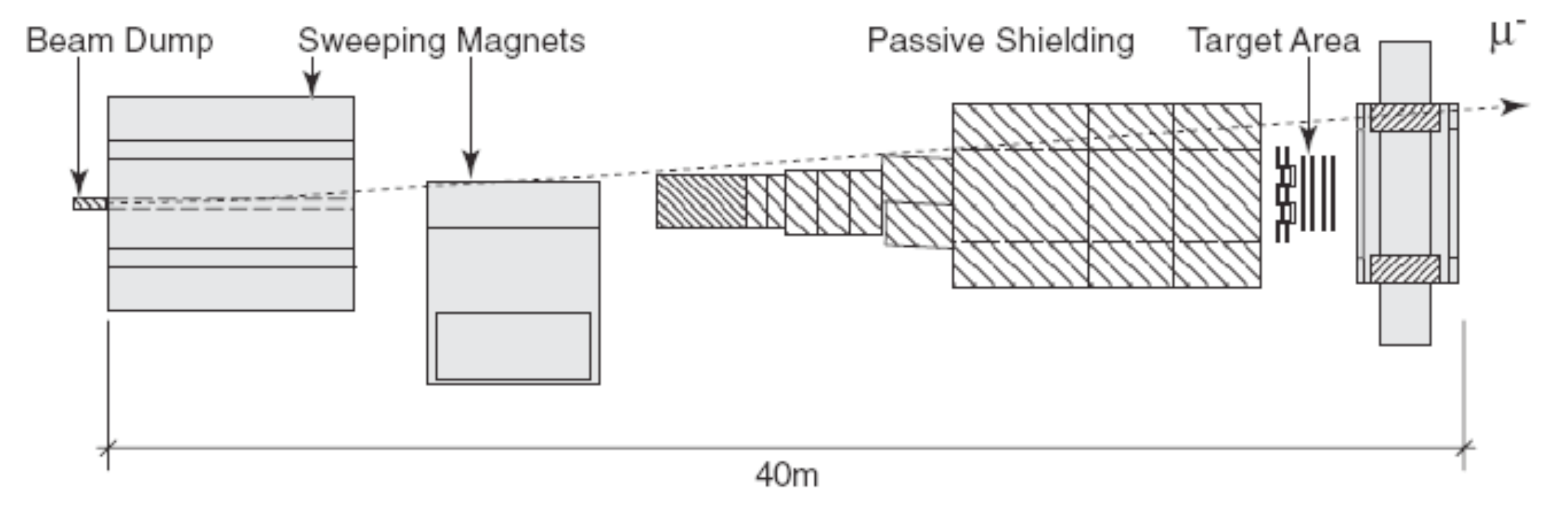}
\caption{Schematic plan view of the DONuT neutrino beam\protect~\cite{Kodama:2008zz}.
The 800~GeV protons are incident on the beam dump from the left. The emulsion
modules are located within the target area, 36~m from the beam dump. The
trajectory of a 400~GeV/$c$ negative muon is shown. The proposed LArTPC
$\nu_\tau$ observation experiment will be assumed for this discussion to use
a similar neutrino production facility with all detector elements downstream
of the passive shielding replaced by a LArTPC.}
\label{fig:nutau_DONuTbeam}
\end{centering}
\end{figure}

\begin{figure}[tb]
\begin{centering}
\includegraphics[width=5in]{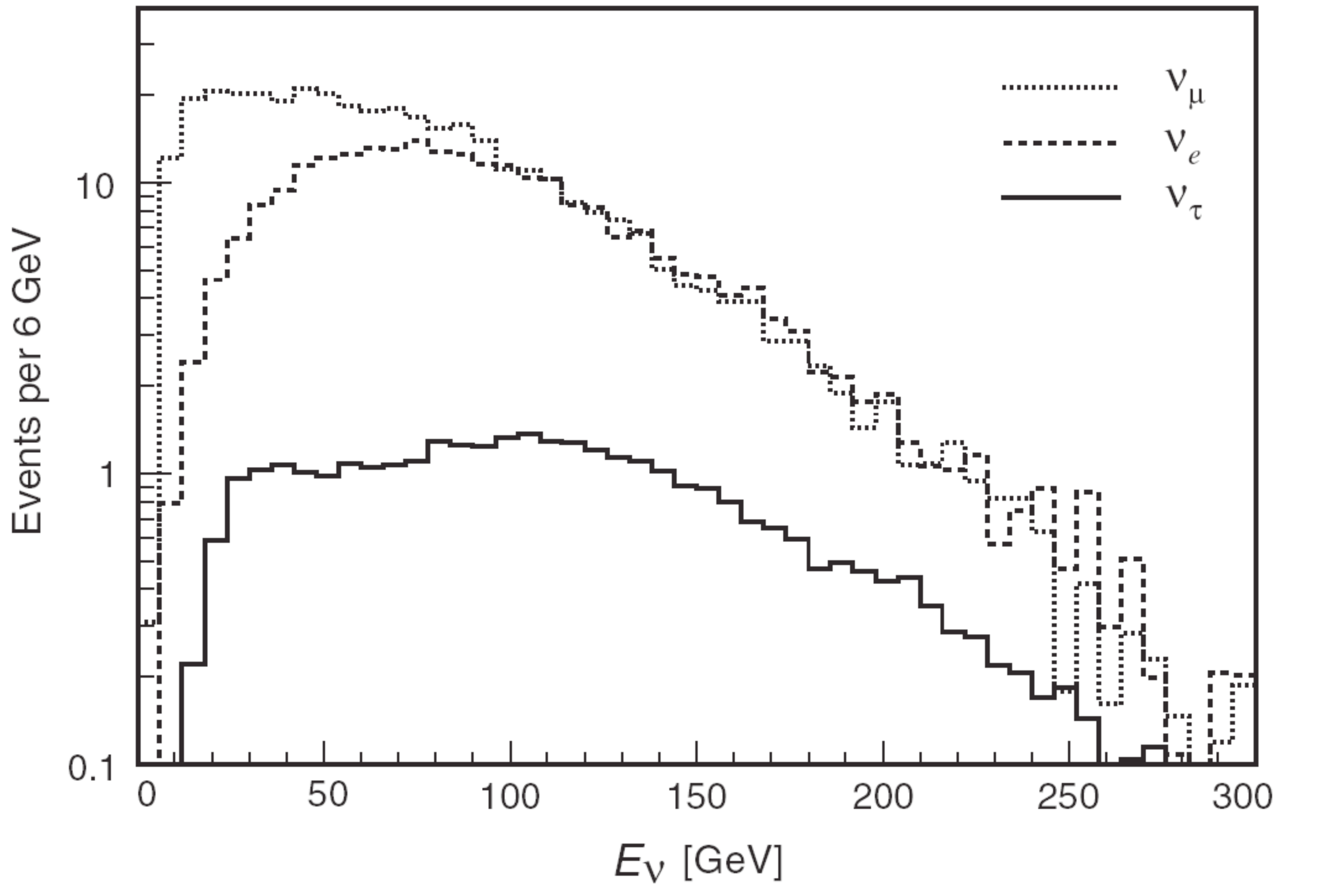}
\caption{Calculated energy spectra of neutrinos interacting in the DONuT
emulsion target\protect~\cite{Kodama:2008zz}. The proposed LArTPC $\nu_\tau$
observation experiment would see similar spectra at a significantly
increased rate.}
\label{fig:nutau_DONuTspectra}
\end{centering}
\end{figure}

For discussion of this $\nu_\tau$ experiment, we assume a similar neutrino
production facility as was used for DONuT~\cite{Kodama:2008zz}, as shown in
Fig.~\ref{fig:nutau_DONuTbeam}, but at the higher intensity described
elsewhere in this document and using a different detector. Neutrinos delivered
to the detector are the result of the decay of particles in hadronic showers
produced by primary proton interactions. The primary proton beam is expected
to be 800~GeV provided by the FNAL Tevatron or CERN SPS+, in which the maximum
center-of-mass energy of an incident proton with a nucleon in the target is
approximately 40~GeV, well above threshold to produce charm as well as bottom
hadrons. Alternatively, if the primary proton beam were produced by a facility
such as CERN's LHC with up to 7~TeV protons in a fixed-target program, the
maximum center-of-mass energy rises to approximately 120~GeV, significantly
enhancing the $\nu_\tau$ flux by the decay of produced on-mass-shell Z$^0$ and
W$^\pm$ bosons to charm hadrons, $\tau^\pm$, and $\nu_\tau$.

After the interaction of 800~GeV protons with the beam dump, $\nu_\tau$ are
produced primarily by the subsequent decay of produced $D_s$ mesons, with a
branching fraction ${\cal B}(D_s^- \to \tau^- \bar\nu_\tau) =
(6.6 \pm 0.6)$\%~\cite{Amsler:2008zzb}. Incident protons are stopped in a beam
dump in the form of a tungsten alloy block; DONuT used a 10~cm $\times$ 10~cm
$\times$ 1~m water-cooled block. The increased intensity of today's
proton facilities may require optimization of the beam dump. Following the beam
dump are dipole magnets sufficient to absorb interaction products and deflect
away high energy muons from the beam center. After the magnets, a passive
absorber is required to further reduce the flux of muons and other interaction
products from the beam center. DONuT used 18 m of steel not more than 2 m from
the beam center for this purpose. Emerging from this absorber are a reduced
flux of muons and a flux of neutrinos of which 3\% will be
$\nu_\tau + \bar\nu_\tau$. The prediction for the spectra of all three
neutrino flavors observed at the DONuT emulsion target and using the DONuT
beam is shown in Fig.~\ref{fig:nutau_DONuTspectra}. The intensity of the
present Tevatron will result in an integrated proton flux approximately 150
times that delivered to DONuT.

\subsection{The detector}

The requirements for an optimal neutrino detector include (a) large mass, (b)
low unit cost, (c) long-term reliable operation, (d) low energy threshold, (e)
high spatial resolution, (f) good energy resolution, (g) homogeneous media
allowing consistent detection capability throughout, (h) density and
radiation length balanced between event containment and spatial resolution of
electromagnetic showers, and (i) high particle-identification efficiency. The
DONuT experiment used a primary detector composed of 260 kg of nuclear
emulsion modules stacked along the beam line, with each module exposed only
for a limited time to avoid track density higher than $10^5$ per cm$^2$. With
the increased intensity of the expected proton beam and significantly larger
event sample required for a precision $\nu_\tau$ appearance measurement, an
emulsion detector is much more difficult. Technologies that may satisfy the
above requirements are water Cherenkov detectors, as employed by ({\it e.g.})
T2K~\cite{Itow:2001ee}, or LArTPCs used by current and developing experiments
ArgoNeuT~\cite{ArgoNeuThttp}, MicroBooNE~\cite{MicroBooNEprop}, and
ICARUS-T600~\cite{Amerio:2004ze}. Spatial and energy resolution, low energy
threshold, and high-efficiency particle identification are characteristics
of LArTPC detectors which will allow the full reconstruction of
$\nu_\tau$ charged-current interactions with efficient identification of the
resulting charged $\tau$. The use of a LArTPC as the primary detector technology
facilitates the identification of typical charged $\tau$ decay products with
excellent vertex and energy reconstruction sufficient to kinematically
reconstruct the intermediate $\tau$. Although $\nu_\tau$ events can be
identified kinematically, it is interesting to note the possibility of
reconstructing the short $\tau$ track in the highest energy interactions
({\it i.e.} a 200~GeV $\tau$ travels a mean distance of 9.7 mm, much larger
than the position resolution along the beam direction). With
this technology, the energy resolution of hits and reconstructed objects
within the detector will allow efficient identification of charged particles
(electrons, muons, protons, pions, kaons) as well as $\pi^0$'s, all necessary
for kinematic reconstruction of charged $\tau$'s. Kinematic separation of
$\nu_\tau$ charged current interactions with $\tau \to \ell \nu \bar\nu$
decays from $\nu_\mu$ and $\nu_e$ charged current interactions is possible by
analysis of missing transverse momentum, non-zero for $\nu_\tau$ charged
current interactions and close to zero otherwise.

\begin{figure}[tb]
\begin{centering}
\begin{tabular}{cc}
\includegraphics[height=2.55in,angle=-90]{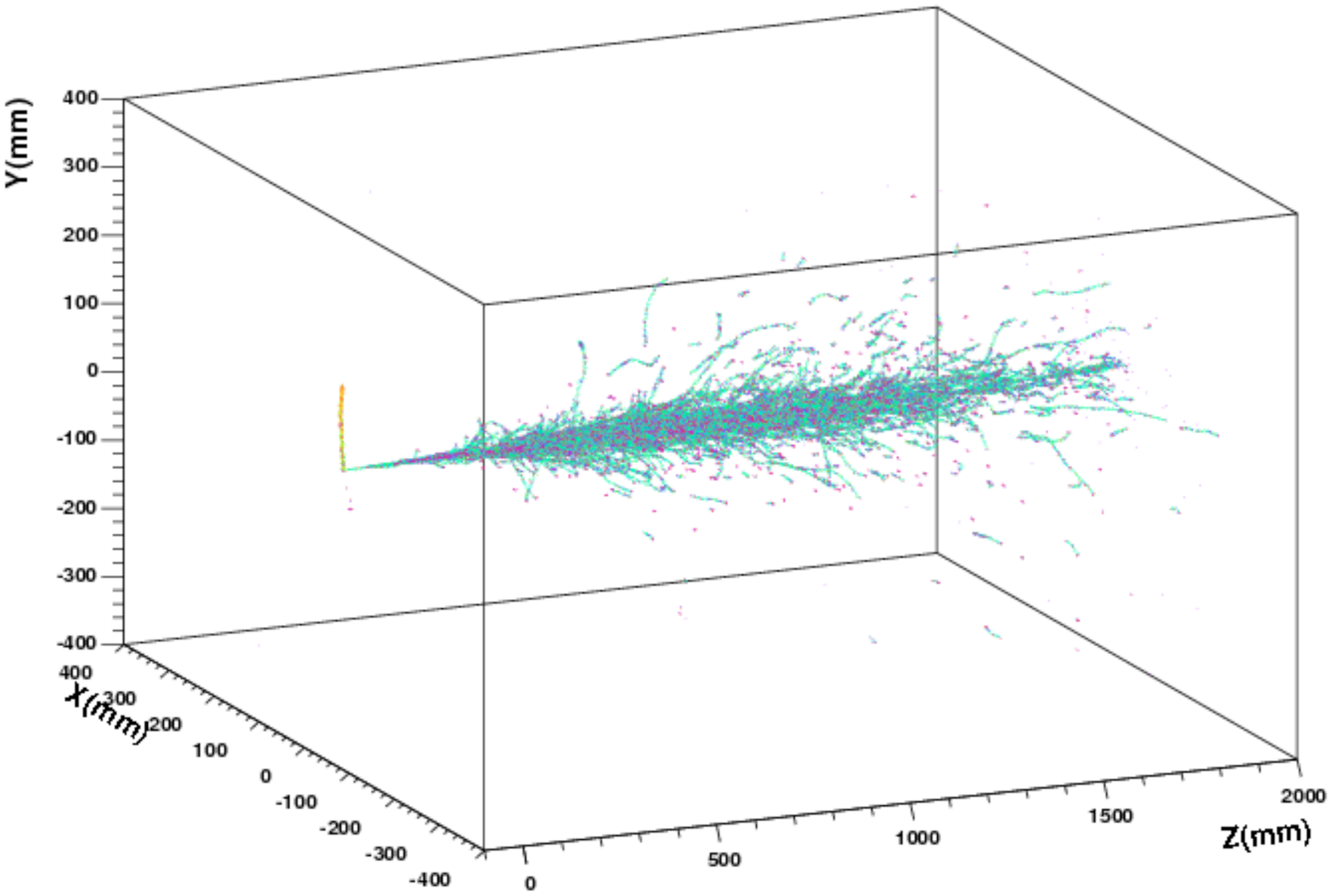} &
\hspace{-1cm}
\includegraphics[height=2.55in,angle=-90]{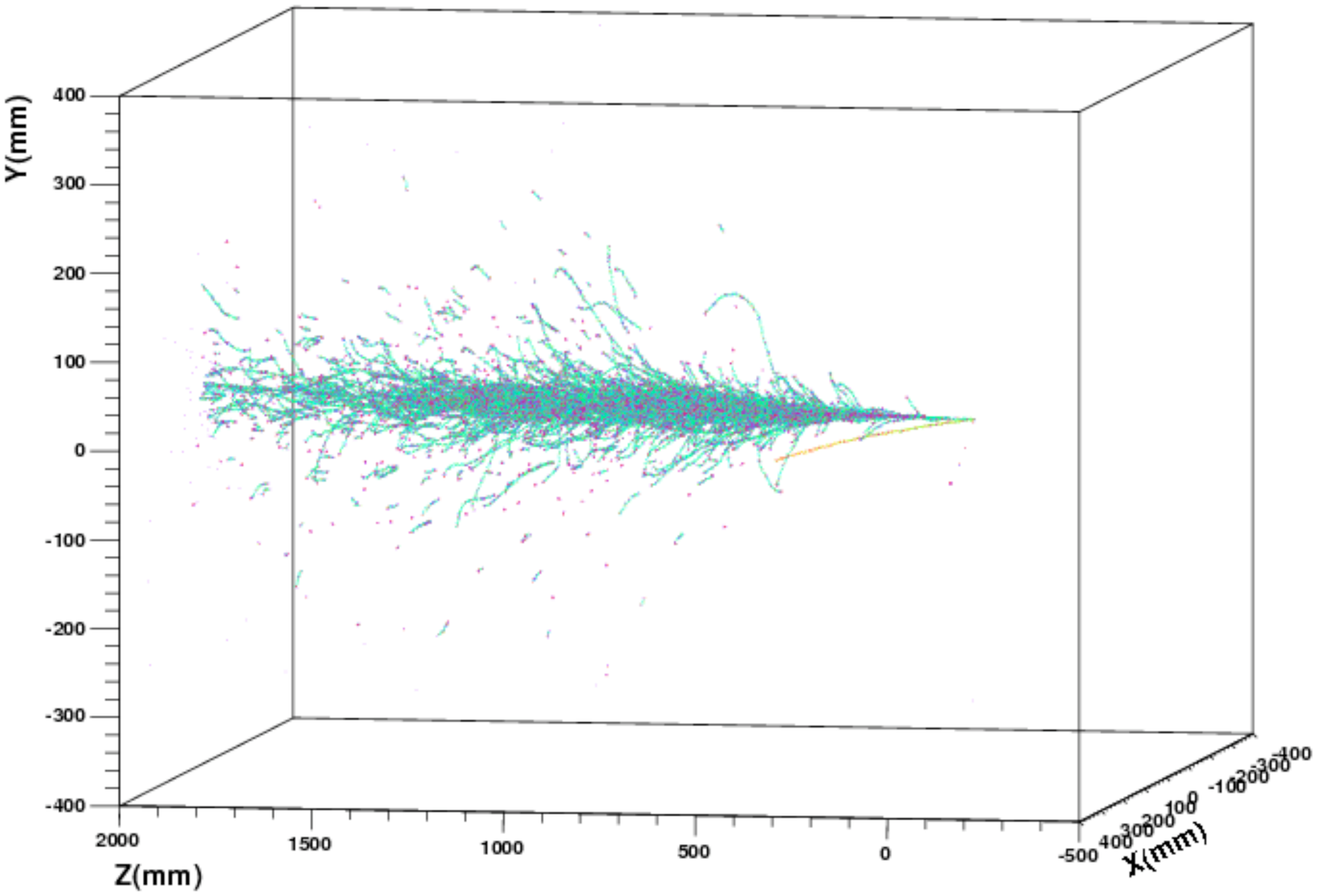} 
\end{tabular}
\caption{Two arbitrary views of a simulated 193~GeV $\nu_\tau$ CC interaction
in a magnetized MicroBooNE-like LArTPC with 3 mm wire spacing in the charge
collection planes. The subsequent 192~GeV $\tau^-$ promptly decays as
$\tau^- \to e^-\nu_\tau \bar{\nu_e}$.  The proton from a recoil resonance
decay ($\Delta^{++} \to p\ \pi^+$) is also clearly visible. The charge drift
direction is along the Y-axis. Assumed single hit position resolution is 3 mm
in X and Z, 1.5 mm in Y.}
\label{fig:nutau_TauENuNu_evd}
\end{centering}
\end{figure}

\begin{figure}[tb]
\begin{centering}
\begin{tabular}{cc}
\includegraphics[height=2.55in,angle=-90]{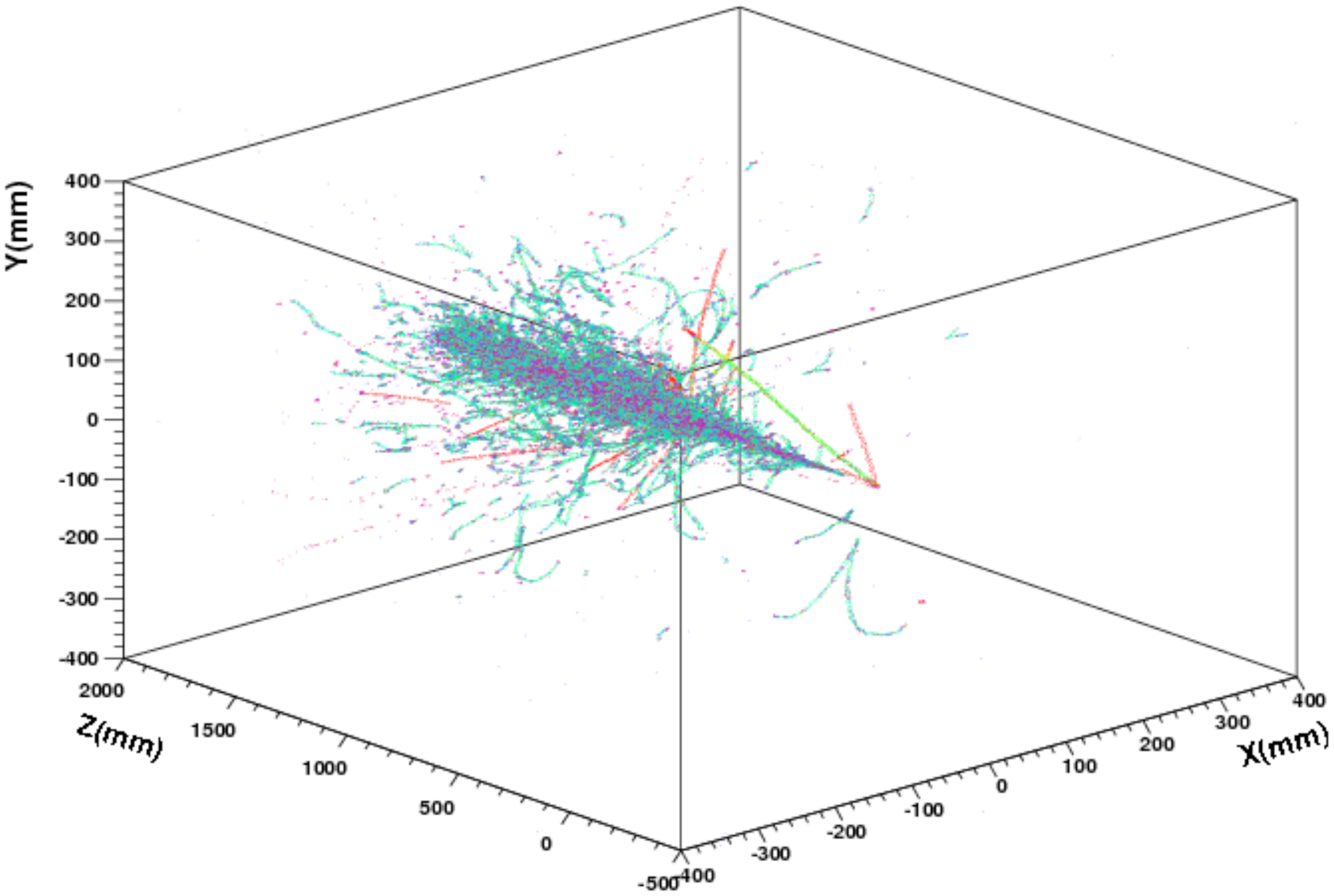} &
\hspace{-1cm}
\includegraphics[height=2.55in,angle=-90]{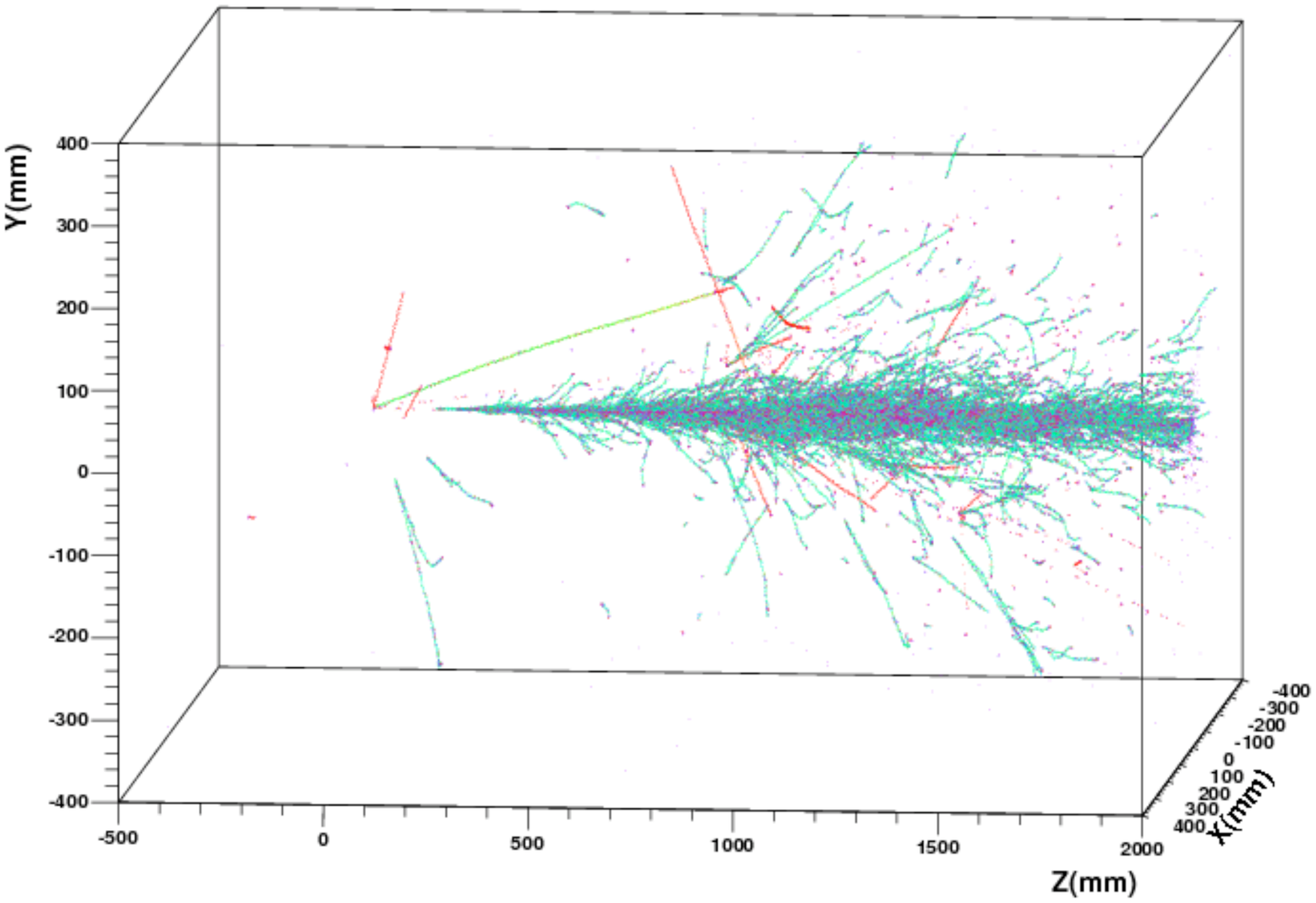} 
\end{tabular}
\caption{Two arbitrary views of a simulated 220~GeV $\nu_\tau$ CC interaction
in a magnetized MicroBooNE-like LArTPC with 3 mm wire spacing in the charge
collection planes. The resulting 212~GeV $\tau^-$  decays as
$\tau^- \to \rho^- \nu_\tau$. The $\rho^-$ daughter $\pi^-$ and $\pi^0$ yield
a clear track and electromagnetic shower emanating from the interaction vertex.
The products of a recoil resonance decay ($\Delta^+ \to p\ \pi^0$) are also
evident as a low energy vertex proton track and two low energy displaced-vertex
daughter photons. The charge drift direction is along the
Y-axis. Assumed single hit position resolution is 3 mm in X and Z, 1.5
mm in Y.}
\label{fig:nutau_TauRhoNu_evd}
\end{centering}
\end{figure}

Preliminary scanning of simulated NC and CC $\nu_e$, $\nu_\mu$, and $\nu_\tau$
events up to 300~GeV, based on a MicroBooNE-like LArTPC with 3mm wire spacing,
verifies the viability of the technology to meet the physics goals of the
proposed experiment. Distinct electromagnetic showers from electrons and
photons, tracks from muons, pions, kaons, protons, as well as displaced
vertices due to ({\it e.g.}) photon conversion, $K_S^0$, and $\Lambda^0$, are
clearly evident, facilitating the identification of individual particles and
resonances. Example simulated $\nu_\tau$ CC events are shown in
Figs.~\ref{fig:nutau_TauENuNu_evd} and~\ref{fig:nutau_TauRhoNu_evd},
demonstrating the expected resolution and pattern of energy deposition hits
from typical interactions. This position and calorimetric resolution
is absolutely necessary in order to kinematically reconstruct a
charged $\tau$ from its decay products.

The concept of adding a magnetic field to a LArTPC has recently been proposed,
with bench tests of its practicality performed on small
detectors~\cite{Badertscher:2004py}. This project may offer the first step at
deploying the technology at the kiloton scale. If a magnetic field is employed
within the TPC, charge-sign identification is possible, which will reduce the
combinatorial background in kinematically reconstructing $\nu_\tau$ charged
current interactions, allow separate $\nu_\tau$ and $\bar\nu_\tau$ measurement,
provide a method of observing potential direct \cpv\ in neutrino
and charged $\tau$ interactions, and provide a second method of track
momentum/energy determination especially useful for events with exiting tracks.

One of the major concerns with LArTPC technology is LAr purity, with current
technology limiting charge drift distance to less than 2-3~m. The neutrino
events at the ${\cal O}$(100~GeV) scale will be very forward boosted, such
that the LArTPC's drift distance can be short relative to the
beam-coordinate. Therefore, sufficient LAr purity may be attained with
existing purification technology, even for a kiloton-scale detector, at the
expense of additional readout planes or a smaller modular geometry.

Combining the increased flux of protons delivered by the proton source ({\it e.g.}
the FNAL Tevatron) with the use of a LArTPC
detector with a mass $2 \times 10^3$ times that of DONuT's 0.5~ton as well as
twice the running time, will result in a delivered flux observed at the
detector approximately $6 \times 10^5$ that observed by DONuT over its
six-month run, equivalently $\cal O$(6 million) $\nu_\tau$ charged current
interactions with one year of data.

\subsection{The Standard Model and beyond}

Here we highlight the prospects for measuring charged and neutral current
$\nu_{\tau}$--matter scattering, observing $\nu_{\tau}$--electron scattering
and probing electromagnetic properties of the tau neutrino. In the Standard
Model, $\nu_{\tau}$ charged current interactions are mediated by $W$-boson
exchange. There is only one measurement (with error bars around 50\%) of the
charged-current scattering cross-section with initial-state tau
neutrinos~\cite{Kodama:2008zz}, and it agrees with Standard Model expectations.
The expectations are that the $\nu_\tau\to\tau$ transitions are
well-described by the Standard Model thanks to abundant data on $\tau$ lepton
processes, including $\tau\to\nu_{\tau}\ell\nu_{\ell}$ ($\ell=e,\mu$),
$\tau\to \nu_{\tau}$+hadrons, $D_{(s)}\to\tau\nu_{\tau}$, etc. The
measurement precision of $\nu_\tau$ charged-current events is of the
utmost importance as it provides a normalization for neutral-current
measurements, which are only very poorly constrained. Furthermore, a
$\tau$-lepton sensitive neutrino detector may also place bounds on
flavor-violating processes such as $\nu_{e,\mu}+X\to\tau+Y$. Even
though these are already strongly constrained by the NOMAD
experiment~\cite{Astier:2001yj}, a $\nu_\tau$-rich beam might
significantly improve on current bounds.

In the Standard Model, neutral current interactions are mediated by $Z$ boson
exchange. Unlike charged-current processes, neutral current processes
involving $\nu_{\tau}$ are only very poorly constrained, especially for
interactions with final state $\nu_{\tau}$ and $\nu_e$~\cite{Davidson:2003ha}.
In more detail, if we add to the Standard Model effective operators of the
type (see Eq.~(\ref{eq:leff}))
\begin{equation}
\mathcal{L}_\mathrm{NSI}^{\nu_{\tau}} =
\sum_{f=e,u,d}\frac{\sqrt{2}}{\Lambda^2}
\Bigl[\, \bar{\nu}_\alpha \gamma_\sigma P_L \nu_\tau 
\,\Bigr]
\Bigl[\,
 \cos\theta_f\, \bar{f}\gamma^\sigma P_L f
+\sin\theta_f\, \bar{f}\gamma^\sigma P_R f
\,\Bigr]  \;,
\label{eq:leff_tau}
\end{equation}
current data constrain $\Lambda\lesssim 100$~GeV for all $f$ and $\theta_{f}$
for $\alpha=e,\tau$. If present, such weak-scale new physics processes
are not only allowed but known to significantly impact the interpretation of
neutrino oscillation experiments (see, for example, Ref.~\cite{Friedland:2005vy} for a detailed discussion). A high statistics
$\nu_\tau$-rich experiment should be able to significantly improve on
current bounds or, perhaps, reveal new physics in the neutrino sector.

Finally, a high statistics experiment should also be sensitive to
$\nu_{\alpha}+e$-scattering events. These can be used (see section on
NuSOnG) to look for different manifestations of physics beyond the
Standard Model. With a $\nu_{\tau}$-rich beam, one can place bounds on
what is usually referred to as the magnetic moment of the tau
neutrino. In more detail, one is sensitive to interactions of the type
\begin{equation}
\mathcal{L}_{\rm mag. mom.} = \frac{\lambda^{\alpha\beta}}{\Lambda}
\Bigl[\, \bar{\nu}_\alpha \sigma^{\rho\sigma} \nu_\beta 
\,\Bigr]F_{\rho\sigma},
\label{eq:mu_tau}
\end{equation}
where $\alpha,\beta=e,\mu,\tau$ and $F_{\rho\sigma}$ is the electromagnetic
field-strength. The nature of the dimensionless coefficients $\lambda$ depends
on the nature of the neutrino fields (Majorana versus Dirac) and their
magnitude is expected to be negligibly small in the absence of new physics
beyond the Standard Model, which here is characterized by the new physics
scale $\Lambda$. What is referred to as the magnetic moment of a particular
neutrino flavor is process dependent and involves different functions of the
$\lambda$ coefficients. While bounds on the $\nu_e$ and $\nu_\mu$
magnetic moments are presently many orders-of-magnitude better than
that of $\nu_\tau$, it is clear that the information one can acquire
with a next-generation $\nu_\tau$ experiment is independent from and
possibly competitive with measurements previous obtained with $\nu_e$
and $\nu_\mu$ scattering (even if neutrinos are Majorana fermions and
the appropriate matrix of $\lambda$-coefficients is
anti-symmetric). The proposed $\nu_\tau$ experiment is wholly
complementary to a next-generation program measuring $\nu_e$ and
$\nu_\mu$ scattering, such as NuSOnG. Astrophysics also provides some
stringent flavor-independent bounds, but these are often model
dependent and need to be confirmed by terrestrial
experiments. Finally, we note that other electromagnetic properties of
the tau neutrino can be probed by neutrino electron scattering (see,
for example, Ref.~\cite{Hirsch:2002uv}).

Additionally, an intense $\nu_\tau$-rich neutrino beam offers a potentially
large sample of highly polarized single charged $\tau$ leptons which may be
uniquely exploited to measure the charged $\tau$ anomalous magnetic moment
form factor as well as the \cpv\ electric dipole moment. This sample of
neutrino-produced single $\tau$'s may also provide an independent measurement
of other $\tau$ properties in an environment with very different systematic
uncertainties than those of the electron-positron collider experiments where
the vast majority of $\tau$ physics has been studied in the last three decades.

\subsection{Primary measurements}

The combination of  LArTPC's fast triggering, high spatial and energy
resolution, and particle identification by specific ionization energy
loss allows a rich program of neutrino physics. The primary
measurement of this experiment will be the high precision relative
cross section measurement, $\sigma_\tau/\sigma_{(\mu,e)}$, for charged
current interactions of $\nu_e$, $\nu_\mu$, and $\nu_\tau$ neutrinos,
which provides a sensitive test of the Standard Model as outlined in
the previous section. When combined with measurements/limits from
NuSOnG or current limits on the magnetic moment of $\nu_\mu$ and
$\nu_e$, similar searches for events consistent with neutrino magnetic
moment interactions in a $\nu_\tau$-rich beam can provide sensitivity
to the $\nu_\tau$ magnetic moment comparable to the present limits for
$\nu_e$ and $\nu_\mu$.

Despite the much smaller sample of $\tau$'s as compared to present
B-factories (${\cal O}(10^9)$ $\tau$'s), the unique environment of this
detector and production mechanism provide a very different set of systematic
uncertainties which allow an interesting laboratory for the verification of
virtually all $\tau$ properties, including branching fraction measurements
as small as ${\cal O}(10^{-5})$. For example, utilizing the sample of
${\cal O}(10^6)$ charged $\tau$ leptons resulting from $\nu_\tau$ charged
current interactions with one year of exposure, several measurements of
charged $\tau$ properties are also possible. In particular, due to the
significant and predictable polarization of the single charged $\tau$'s
produced by $\nu_\tau$ charged current interactions, this experiment is
potentially much more sensitive to the anomalous magnetic moment form factor
and \ electric dipole moment of the charged $\tau$ than previous
experiments.

Further neutrino physics which may be measurable in this detector includes
exclusive cross section measurements, such as coherent-pion production in
neutral current and charged current interactions as well as $\nu_\tau e$
charged and neutral current interactions. The significant size and low energy
threshold of the LArTPC also allows measurement of solar neutrino rates as
well as burst-supernova neutrino sensitivity out of time with the beam spill. The
proximity of the detector to the surface, expected pointing resolution of
reconstructed tracks, and the size of the detector will yield a significant
rate of cosmic-ray induced muons offering a wealth of interesting potential
opportunities ranging from the observation of climactic changes in the
atmosphere to searching for point sources of cosmic rays and sensitivity to
the solar magnetic field.

\subsection{$\nu_\tau$ summary}

Though the $\nu_\tau$ has been assumed to exist for over thirty years,
only nine $\nu_\tau$ charged current interactions have been observed
directly. Precision study of this particle and its interactions is clearly
warranted in order to determine if its nature is as predicted by the
Standard Model and provides a unique laboratory to search for new physics.
The proposed program of neutrino and charged $\tau$ physics is broad enough
to support a wide variety of studies alongside the primary studies of the
electroweak interactions of the $\nu_\tau$ using ${\cal O}(10^5)$ larger
sample of interactions than has been observed previously. The proposed 1~kt
size of the LArTPC for this experiment is a natural choice to achieve the
desired physics goals while also providing an intermediate step in the
development of the LArTPC technology between MicroBooNE (100 t) and DUSEL
(5+ kt). The addition of a solenoidal field would significantly enhance the
physics capabilities of the project while pioneering the technological
advancement of coupling LArTPCs with a magnetic field at the kiloton scale.

\section{Searches for Exotic Neutrinos}

Singlet (sterile) neutrino states arise in models which try to
implement massive (light) neutrinos in extensions of the Standard
Model. Three singlet states $N_1$, $N_2$ and $N_3$ are associated with the
three active neutrinos. In the original see-saw mechanism, these new
states have very large masses, but variations like the nMSM model~\cite{Asaka}
give them masses which are within reach of experimental searches.
Limits exist from laboratory experiments, but they extend to masses up
to 450~MeV, and apply to couplings with the $\nu_e$ or $\nu_\mu$.  An upgraded
Tevatron machine could enlarge the domain of exploration in masses and
couplings with the study of neutrinos coming from $D$ and $B$ decays. For
the first time, mixings to the $\nu_\tau$ could be efficiently
investigated. Such a search can be envisaged in the beam-dump of the
$\nu_\tau$ experiment.

\subsection{Production of sterile neutrinos}

If heavy neutrinos exist, they mix with active neutrinos through a
unitary transformation. Any neutrino beam will contain a fraction of
heavy neutrinos at the level $U_{Nl}^2$ where $U$ denotes the mixing
matrix element between the heavy state $N$ and the charged lepton, $l$
($l$ being $e$ or $\mu$ or $\tau$).  At low energy accelerators,
neutrinos are produced in $\pi$ and K decays. At higher energies,
charm and beauty contribute. Kinematically, the mass range allowed for
the production of a heavy $N$ depends on the emission process. In $\pi
\rightarrow \mu + N$ decays, sterile neutrinos can reach a mass of 30~MeV. In $\pi \rightarrow e+N$ channels, the range increases to 130~MeV. Kaons allow larger masses, up to 450~MeV. $D$ decays extend the
range to $\sim$1.4~GeV for $e$ and $\mu$ channels, (but only to 180~MeV
for the $\tau$ channel), and $B$ decays to $\sim$4.5~GeV (3~GeV for the $\tau$
channel).  The flux of N accompanies the flux of known neutrinos at
the level of $U_{Nl}^2$. Corrections to this straightforward result
come from helicity conservation which applies differently here. For
example, for massless neutrinos, it suppresses $\pi \rightarrow e+\nu$
decays relative to $\pi \rightarrow \mu + \nu$ decays. This is not
true anymore for $\pi \rightarrow e+N$. Phase space considerations
have also to be taken into account. Thus, precise calculations have to
be done in all possible cases to be considered. For example, precise branching fractions in the case of massive neutrinos have been calculated in Ref.~\cite{Levy}.

\subsection{Decays of sterile neutrinos}

$N$'s are not stable. They will decay through purely weak
interactions. The lifetime critically depends on the mass considered;
it varies as $m^5$ power. Decay modes also depend on the $N$ mass. As
soon as the mass is greater than 1~MeV, the first channel to open is
$N\rightarrow ee \nu$. With increasing masses, new modes open, and one can obtain
$e\mu\nu$, $\pi e$, $\mu \mu \nu$, $\pi\mu$. For higher mass states
potentially produced in B decays, new modes become relevant. For
example, for masses above 2~GeV, one can envisage the channels $De$, 
$D\mu$
or even $D\tau$. Exact branching fractions require precise calculations.
The lifetime is given by the formula applying to weak decays, apart
from a general suppression factor coming again from the mixing
$U_{Nl}^2$. Other factors coming from helicity and phase space
considerations have to be included.

\subsection{Previous results}

The search consists in looking for a decay signature, typically two
charged tracks, one of them being a lepton, and reconstructing a vertex in
an empty volume. If no candidates are found, one sets a limit in a
two-dimensional plane, mass vs. mixings. Mixings can be equal or different
in production and decay. Thus, one tests 6 different
combinations of mixings in principle.  This has been attempted at CERN by the low
energy experiment PS191~\cite{Bernardi} with $5\times10^{18}$ protons of 19~GeV
on target, or about $10^{15}$ neutrinos (essentially all $\nu_\mu$) crossing
an empty detector volume. The neutrinos were produced in $\pi$ and $K$
decays. Thus the limits apply to couplings to $\nu_e$ and
$\nu_\mu$. Kinematically, the $\tau$ is not accessible either in production or
in decay. The explored mass range is limited to the $K$ mass.  The
limits on the $U_{Nl}^2$ couplings reach the level of $10^{-8}$ in a large range of
accessible masses and for all combinations of mixings to $e$ or
$\mu$. Soon-to-run experiments (such as MINERvA~\cite{minerva}) could improve these results by an order of
magnitude.  In order to increase the domain of exploration, it is
necessary to consider higher energy beams producing neutrinos via $D$
and $B$ decays.  $D_s$ decays into $\tau \nu_\tau$, with a branching fraction of 6\%.
$B$’s decay into the 3 leptonic channels, $X e \nu_e$, $X \mu \nu_\mu$, 
$X \tau \nu_\tau$, with
branching fractions 10\%, 10\% and 5\%, respectively. This allows the
search of $N$ states with masses up to 4.5~GeV, mixing in particular
with the $\nu_\tau$.  Since the limits vary as the square root of the
accumulated neutrino flux, the number of protons on target has to be maximal.

\subsection{Detector considerations}

\begin{figure}[htb]
\centering
\scalebox{0.5}{\includegraphics[clip=true]{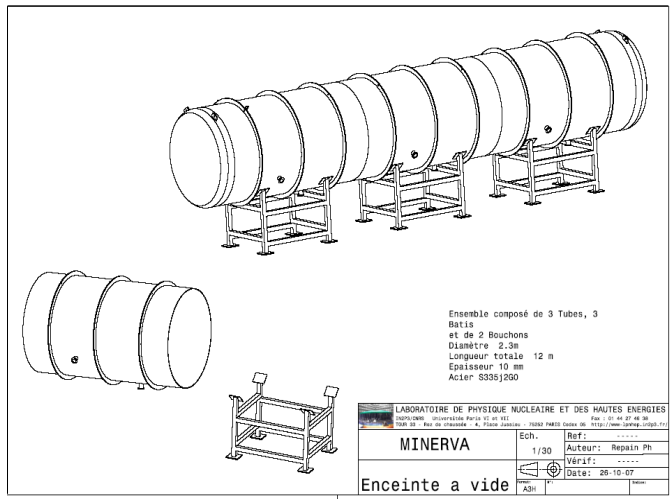}}
\vspace{-6.3cm}
\caption{A 12 m long tank for the evacuated decay volume.}
\label{fig:decayvolume}
\end{figure}

The experiment consists in detecting a decay vertex arising in an
empty volume set in a neutrino beam and characterized by, in most
cases, two charged tracks.  The detector requires a decay volume as
large as possible followed by a calorimeter. In principle, the search
is better done at low energy. However, in order to extend the region of
potential masses, one has to produce B’s and this is only done at high
energy.  The advantage of an upgraded Tevatron machine comes directly
from the much increased luminosity available. If, instead of being done in  a beam dump, the search is done in a neutrino beam, for example in  parallel with NuSOnG,
the background coming from neutrino interactions is also substantially
increased. In 12 m of air the number of interactions amounts to
several 10000 events. Charged current events will give a muon in the final
state in 99\% of the cases. It becomes essential to have an evacuated
volume with a calorimeter to be able to efficiently identify
electrons and muons.  Studies have been made for the decay volume. 
A 12 m long pipe where the vacuum can be pushed down to
10$^{-3}$~atm can be seen in Fig.~\ref{fig:decayvolume}. The background from interactions becomes manageable. A
higher vacuum would require much more sophisticated techniques.  A
good spatial resolution plane is necessary between the decay volume
and the calorimeter in order to precisely reconstruct the decay
vertex.  The best limits on couplings come from exclusive channels: $\pi e$
and $\pi \mu$ for low masses, $K e$ and $K \mu$ for intermediate masses, and $D e$, $D \mu$, $\pi \tau$ and $D \tau$ for higher masses. The decay channels involving $e$ and $\mu$
can be totally reconstructed. Two essential constraints arise: 1) the
reconstructed direction of arrival must point to the neutrino
production target and 2) the invariant mass of the detected particles must
reconstruct a fixed mass.  For example, one can search for a  $D + e \rightarrow K + \pi + \pi + e$. This means that the calorimeter must have
the track reconstruction and identification capability. It must be
fine grained and preferably come with a magnetic field.  These constraints are
not applicable for decay modes involving a $\tau$ lepton where a
characteristic $\pi \tau$ will show up.

\subsection{Expectations}

Extrapolating the neutrino fluxes used in the $\nu_\tau$ experiment, one
expects about 10$^{16}$ $\nu_\tau$ per year traversing a 3 m$^2$ section with average energy of 50~GeV. These neutrinos come from  $D_s$ decays. Other
$D$’s give about 20 times more $\nu_e$ and $\nu_\mu$. With a ratio of production
cross-sections $B/D \sim 10^{-3}$ one expects of the order of $10^{14}$ neutrinos
of each type coming from $B$’s.  With these numbers, one can estimate the
limits obtained by a null experiment in a 10 m long decay volume. 
\begin{itemize}
\item From $D$ decays one reaches a $U^2$ limit of $10^{-9}$ for a mass around 1~GeV
and mixings to $e$ and $\mu$.  
\item From $B$ decays one can reach $10^{-8}$ for all mixings, in particular the never explored $U^2_{\tau \tau}$ around a mass of 3~GeV.
\end{itemize}

Heavy neutral leptons arise in models which try to accommodate massive
active neutrinos. Searches have been done in low energy neutrino
beams. The advantage of an upgraded high energy machine is two-fold:
the high energy allows exploration in a larger domain of masses, up to the
$B$ mass, and the high luminosity pushes down the limits. In
particular, it can set meaningful limits on the practically unexplored
couplings to the $\tau$.  The fascinating possibility of finding sterile
neutrinos could be uniquely tested in such an experiment.

\section{Conclusion}

This paper presents examples of the compelling physics that can result
from a $\sim$1~TeV fixed target facility.    We have especially
highlighted the discovery potential in the charm sector, which would
utilize slow-spill beams.  We also considered forefront physics in the
neutrino sector.  We reviewed an existing idea for precision
electroweak studies.  Also, we presented two new promising and unique
avenues for beyond standard model neutrino searches using beam dump
production.  The first of these uses $\nu_\tau$ charged current events
which are produced above threshold by a proton beam in the 800~GeV to
1~TeV range.  The second is a search for neutral heavy leptons
produced in the beam dump.  

This combination of experiments represents
an integrated program aimed at discovery of new physics which 
is complementary to other approaches under discussion for the future.

\newpage

\appendix

\section{Specifics of Running 1~TeV Beams: The Tevatron}

Previous 800~GeV fixed-target operation of the Tevatron ran with a
maximum throughput of roughly 25-28 $ \times 10^{ 12}$ protons
(25-28~Tp) per pulse every 60~sec with a duty cycle of roughly
33-40\%.  
The beam was shared, over a 20-23~sec flat-top period, between slow
spill experiments and neutrino experiments which required fast
extracted beams. To meet the demands of NuSOnG, the facility needs to
be able to deliver approximately $2\times 10^{20}$ protons on target
over five years of running at 66\% overall operation efficiency per
year.  This translates to an average particle delivery rate during
running of about 1.8~Tp/sec.  Assuming that only a 40~sec ramp will
be required for NuSOnG, each ramping of the Tevatron would need
to deliver about 75~Tp, more than 2.5~times the previous record
intensity.  The subsections below address some of the major issues
regarding re-institution of a Tevatron fixed-target program, and
issues associated with meeting the above intensity demand.

\subsection{Magnet ramping}

The original Tevatron fixed-target program ran at 800~GeV and stress
and strain on the superconducting magnets was a major issue early in
the program.  Issues with lead restraints within the cryostat were
eventually identified and all dipole magnets were repaired in the
tunnel in the late 1980's.  Since that time, the Tevatron has been
able to average over 250,000 cycles between failures of dipole
magnets~\cite{Annala}.  This
``rate'' includes failures of collider-specific magnets, such as
low-beta quadrupoles.  Note that a neutrino program which demands
$2\times 10^{20}$~POT, using a synchrotron that delivers 75~Tp every
cycle, requires about 2.7 million cycles -- thus, on the order of 10
failures could be expected during the course of the experiment.

Once the fixed-target operation was halted and only collider
operation was foreseen, the capability to repair and rebuild Tevatron
magnets was greatly reduced at the laboratory.  However, assuming no
need for building new magnets from scratch, capabilities still exist
to perform repairs and, along with the given inventory of spare
Tevatron magnets and corrector packages, a multi-year fixed-target
operation consistent with the above is sustainable from this
aspect~\cite{Harding}.

Ramp rate studies of Tevatron dipole magnets have been performed, and
rates of 200-300~A/sec can be maintained at 4.6$^\circ$~K without
quenching~\cite{MTF}.  The
current power supply system can still perform at this level.  To
increase reliability, however, some PS system components may need to
be upgraded.  Additionally, the Tevatron RF system is still capable of
running in the fixed-target state, though beam loading effects and
appropriate compensation will need to be investigated for the
anticipated higher intensity operation.  Two Main Injector (MI) pulses
would be used to fill the Tevatron.  At 3~sec per 150~GeV MI cycle,
this constitutes a 15\% impact on other MI demands.

\subsection{Comments on high intensity}

The record intensity extracted from the Tevatron in a cycle at 800 GeV
was almost 30~Tp in 1997, though 20-25~Tp was far more typical.  At
that time, the bunch length during acceleration would shrink to the
point where a longitudinal instability at higher energies ($\sim$600
GeV), resulting in aborts and sometimes quenches.  This was
compensated as well as possible with ``bunch spreading'' techniques
(blowing up the emittance via RF noise sources).  Today, the Main
Injector is capable of providing greater than 40~Tp per pulse, which
could, in principle, fill the Tevatron to 80~Tp.  Many improvements to
the Tevatron beam impedance have been made during Run II, including,
for example, reduction of the Lambertson magnet transverse impedances
which were identified as major sources.  Additionally, advances in RF
techniques/technology and damper systems, {\em etc}., may allow, with
enough studies and money, much better compensation of these effects,
if required.  This is a primary R\&D point, if intensities near 75~Tp
are to be realized in the Tevatron.

\subsection{Re-commissioning of extraction system}

Returning the Tevatron to fixed-target operation would require the
re-installation of the extraction channel in the A0 straight-section
from which beam would be transported to the existing Switchyard area
and on to the experimental target station.  The electrostatic septa
were located at the D0 straight section and could straightforwardly be
reinstalled in the original configuration.  All of this equipment
is currently in storage and available for use.  The B0 straight
section, currently housing the CDF detector, would be replaced with standard long straight section optical components.  Thus, the higher heat leak elements
presently installed in the B0 and D0 regions would be absent, 
requiring less demands from the cryogenics system.

The other necessary piece of hardware is the slow-spill feedback
system, referred to as ``QXR'' which employs fast air-core quadrupoles
installed at warm straight sections in the Tevatron for fast feedback
tune adjustment during the resonant extraction process.  Again, this
equipment mostly still exists, though it may be desirable to perform a
low-cost upgrade to modernize some electronic components.

The neutrino experiment being discussed has requested ``pinged'' beam,
short bursts of particles brought about by the QXR system.  NuSOnG
will likely require tens of {\em pings} per cycle, during an assumed 1
sec flat-top. Resonant extraction is an inherently lossy process, on
the scale of 1-2\%, determined by the particle step size across the
thin electrostatic septum wires.  Historically, loss rates were
tolerable with between 20-30~Tp extracted over 20 sec.  Pings, each lasting on the scale of 1-2~ms with approximately 5~Tp per ping -- and sometimes higher -- were extracted routinely for the Tevatron neutrino program.  Thus, 15 or more such pings over a 1~sec flat top should be straightforward.  Alternative methods for fast extraction could be
contemplated, though perhaps at a price.  For instance, if an
appropriate RF bunching scheme (using a 2.5~MHz RF system, for
example) can be employed to prepare bunches spaced by 400~ns, then a
fast kicker magnet system might be able to extract 50 such bunches
one-by-one to the Switchyard, a much cleaner extraction process.
Spreading the beam across fewer, longer bunches may also help to
mitigate coherent instability issues.  This opens up another possible
R\&D point to pursue.  To set the scale, the highest intensity
extracted in a single pulse ({\it i.e.} not during a slow spill)
without quenching the Tevatron was about 10~Tp~\cite{Annala}.
(Also, this was a test, not a normal
operational procedure.)

The exact method used for 800~GeV operation would be a point closely
negotiated between the laboratory and the experiment(s) using the
beam.  Both resonant extraction and kicker methods should be feasible
within reasonable constraints.

\subsection{Tevatron abort system}

The abort system used during high intensity fixed-target operation was
located at C0 and was capable of absorbing 1~TeV proton beams at
30~Tp, repeatedly every ``several'' seconds, to the abort dump.  While
not used in collider operation, this beam dump and beam delivery
equipment near the C0 straight section is still available and still
accessible, and requires re-installation of extraction devices and
their power supplies.  The ultimate parameters of the neutrino
experiment being discussed pushes the beam stored energy from about
3.5~MJ (27~Tp at 800~GeV) toward 10~MJ.  The design limits of this
system would need to be re-examined, and the implications and
environmental impact of re-establishing this area as the primary abort
must be looked at carefully.

\section*{Acknowledgments}

We gratefully acknowledge support from the Department of Energy, the National Science Foundation, the Consejo Nacional de Ciencia y Tecnolog\'{\i}a and the Universities Research Association (URA) Visiting Scholars at Fermilab Award.


\begin{thebibliography}{99}

\bibitem{sps}
R.~Garoby {\it et al.}, ``Scenarios for Upgrading the LHC Injectors'', 
CERN-AB-2007-007 (2006).

\bibitem{sps2}
R.~Garoby, ``Upgrade Issues for the CERN Accelerator Complex'', 
Proc. of EPAC08, Genoa, Italy (2008).

\bibitem{NuSOnGEW} 
  T.~Adams {\it et al.}  (NuSOnG Collaboration),
  Int.\ J.\ Mod.\ Phys.\  A {\bf 24}, 671 (2009).

\bibitem{NuSOnGQCD} 
T.~Adams {\it et al.} (NuSOnG Collaboration), 
``QCD Precision Measurements and Structure Function Extraction 
at a High Statistics, High Energy Neutrino Scattering Experiment: NuSOnG",  
to be submitted to Int.\ J.\ Mod.\ Phys.\  A.

\bibitem{fnal_fixed-target}
J. A. Appel (Ed.), C. N. Brown (Ed.), P. S. Cooper (Ed.), H. B. White (Ed.),
{\it Symposium in Celebration of the fixed-target program with the
Tevatron}, FERMILAB-CONF-01-386 (2000) arXiv:0008076 [hep-ex].

\bibitem{Schwartz}
A. Schwartz, Nucl.\ Phys.\  B {\bf 187}, 224 (2009).

\bibitem{Buras}
I. Bigi {\it et al.\/}, arXiv:0904.1545 [hep-ph] (2009).

\bibitem{Grossman1}
Y.\ Grossman, Y.\ Nir, G.\ Perez, arXiv:0904.0305 [hep-ph] (2009).

\bibitem{Golowich}
E.~Golowich {\it et al.\/}, arXiv:0903.2830 [hep-ph] (2009).

\bibitem{Grossman2}
K.\ Blum, Y.\ Grossman, Y.\ Nir, G.\ Perez, arXiv:0903.2118 [hep-ph] (2009).

\bibitem{Li}
X. Li and Z. Wei, Phys. Lett B {\bf 651}, 330 (2007).

\bibitem{Ball}
P. Ball, J. Phys. G {\bf 34}, 2199 (2007).

\bibitem{Blanke}
M. Blanke {\it et al.\/}, Phys. Lett. B {\bf 657}, 81 (2007).

\bibitem{Nir} Y. Nir, JHEP {\bf 0705}, 102 (2007).

\bibitem{e791_kpi} E.\ M.\ Aitala {\it et al.\/} (E791 Collaboration),
Phys.\ Rev.~D {\bf 57}, 13 (1998).

\bibitem{e831_kpi} J.\ M.\ Link {\it et al.\/} (FOCUS Collaboration),
Phys.\ Lett.~B {\bf 618}, 23 (2005). 

\bibitem{babar_kpi} B.\ Aubert {\it et al.\/} (BaBar Collaboration),
Phys.\ Rev.\ Lett.\ {\bf 98}, 211802 (2007).

\bibitem{belle_kk} M.\ Staric {\it et al.\/} (Belle Collaboration),
Phys.\ Rev.\ Lett.\ {\bf 98}, 211803 (2007).

\bibitem{cdfD0mix} T.\ Aaltonen {\it et al.\/} (CDF Collaboration), 
Phys.\ Rev.\ Lett.\ {\bf 100}, 121802 (2008).

\bibitem{herab_kpi} I.\ Abt {\it et al.\/} (\hb\ Collaboration),
  Eur.\ Phys.\ Jour.~C {\bf 52}, 531 (2007) .

\bibitem{dcstocf} C.\ Amsler {\it et al.\/} (Particle Data Group), 
Phys.\ Lett.~B {\bf 667}, 1 (2008).

\bibitem{belle_kpi} L.\ Zhang {\it et al.\/} (Belle Collaboration),
Phys.\ Rev.\ Lett.\ {\bf 96}, 151801 (2006).

\bibitem{superbelle} {\tt http://superb.kek.jp/}

\bibitem{superB} {\tt http://www.pi.infn.it/SuperB/}

\bibitem{lhcb_kpi} P.\ Spradlin, G. Wilkinson, F. Xing {\it et al.},
LHCb public note LHCb-2007-049 (2007).

\bibitem{hfag_charm_fits} 
{\tt http://www.slac.stanford.edu/xorg/hfag/charm/FPCP08/results\_mix+cpv.html}

\bibitem{hfag_charm_dcpv} 
E. Barbierio {\it et al.\/} (Heavy Flavor Averaging Group), arXiv:0808.1297 [hep-ex] (2008).

\bibitem{CLEOc2007} S. Dobbs {\it et al.} (CLEOc Collaboration), 
Phys. Rev.~D {\bf 76}, 112001 (2007). 

\bibitem{FOCUS2002} J.~M. Link {\it et al.} (FOCUS Collaboration), 
Phys. Rev. Lett. {\bf 88}, 041602 (2002); 
Err., Phys. Rev. Lett. {\bf 88}, 159903 (2002).

\bibitem{E7911997} E.~M. Aitala {\it et al.} (E791 Collaboration), 
Phys. Lett.~B {\bf 403}, 377 (1997).

\bibitem{CLEOc2008} P. Rubin {\it et al.} (CLEOc Collaboration), 
Phys. Rev.~D {\bf 78}, 072003 (2008).

\bibitem{BaBar2005} B. Aubert {\it et al.} (BaBar Collaboration), 
Phys. Rev.~D {\bf 71}, 091101 (2005).

\bibitem{FOCUS2000} J.~M. Link {\it et al.} (FOCUS Collaboration), 
Phys. Lett.~B {\bf 491}, 232 (2000); 
Err., Phys. Lett.~B {\bf 495}, 443 (2000).

\bibitem{E6871994} P.~L. Frabetti {\it et al.} (E687 Collaboration), 
Phys. Rev.~D {\bf 50}, 2953 (1994).

\bibitem{FOCUS2005} J.~M. Link {\it et al.} (FOCUS Collaboration), 
Phys. Lett.~B {\bf 622}, 239 (2005).

\bibitem{Belle2008_2} M. Staric {\it et al.} (Belle Collaboration), 
Phys. Lett.~B {\bf 670}, 190 (2008).

\bibitem{BaBar2008} B. Aubert {\it et al.} (BaBar Collaboration), 
Phys. Rev. Lett. {\bf 100}, 061803 (2008).

\bibitem{CDF2005} D. Acosta {\it et al.} (CDF Collaboration), 
Phys. Rev. Lett. {\bf 94}, 122001 (2005).
\bibitem{ycp_cleo}
S.~E. Csorna {\it et al.} (CLEO Collaboration),
           Phys. Rev.~D {\bf 65}, 092001 (2002).

\bibitem{E7911998} E.~M. Aitala {\it et al.} (E791 Collaboration), 
Phys. Lett.~B {\bf 421}, 405 (1998).

\bibitem{CLEO2001} G. Bonvicini {\it et al.} (CLEO Collaboration), 
Phys. Rev.~D {\bf 63}, 071101 (2001).
 
\bibitem{CLEO1995} J.~E. Bartelt {\it et al.} (CLEO Collaboration), 
Phys. Rev.~D {\bf 52}, 4860 (1995).

\bibitem{BaBar2008b} B. Aubert {\it et al.} (BaBar Collaboration), 
Phys. Rev.~D {\bf 78}, 051102 (2008).

\bibitem{Belle2008} K. Arinstein {\it et al.} (Belle Collaboration), 
Phys. Lett.~B {\bf 662}, 102 (2008).

\bibitem{CLEO2005} D. Cronin-Hennessy {\it et al.} (CLEO Collaboration), 
Phys. Rev.~D {\bf 72}, 031102 (2005).

\bibitem{CLEO2001a} S. Kopp {\it et al.} (CLEO Collaboration), 
Phys. Rev.~D {\bf 63}, 092001 (2001).

\bibitem{Belle2005} X.~C. Tian {\it et al.} (Belle Collaboration), 
Phys. Rev. Lett. {\bf 95}, 231801 (2005).

\bibitem{CLEO2001b} G. Brandenburg {\it et al.} (CLEO Collaboration), 
Phys. Rev. Lett. {\bf 87}, 071802 (2001).

\bibitem{CLEO2004} D.~M. Asner {\it et al.} (CLEO Collaboration), 
Phys. Rev.~D {\bf 70}, 091101 (2004).

\bibitem{GounarisSakurai68} 
G.J. Gounaris and J. J. Sakurai, Phys. Rev. Lett. {\bf 21}, 244 (1968). 


\bibitem{Flatte76} S.\,M.~Flatt\'{e}, Phys. Lett.~{\bf 63B}, 224 (1976).

\bibitem{Wigner46} E.P. Wigner, Phys. Rev. {\bf 70}, 15 (1946).

\bibitem{WignerEisenbud47} E.P. Wigner and L. Eisenbud, Phys. Rev. {\bf 72}, 29 (1947).

\bibitem{Aitchison72} I. J. R. Aitchison, Nucl. Phys. A {\bf 189}, 417 (1972).

\bibitem{Zemach64} C. Zemach, Phys. Rev. {\bf 133}, B1201 (1964).

\bibitem{Kopp01} S. Kopp {\it et al.\/} (CLEO Collaboration), 
Phys. Rev. D {\bf 63}, 092001 (2001).

\bibitem{Lau07} Y.~P. Lau, Ph.D. thesis, Princeton University, 2006 (unpublished).




\bibitem{PDG08_scalars} 
S. Spanier, N.A. Tornqvist and C. Amsler, ``Note on Scalar Mesons,'' in
  C. Amsler {\it et al.\/} (Particle Data Group), Phys. Lett. B {\bf 667}, 1 (2008). 

\bibitem{PDG08_eta} C. Amsler and A. Masoni, 
``The $\eta(1405)$, $\eta(1475)$, $f_1(1420)$, and $f_1(1510)$,'' in
  C. Amsler {\it et al.\/} (Particle Data Group), Phys. Lett. B {\bf 667}, 1 (2008).

\bibitem{selex02} 
M.~Mattson {\it et al.\/} (SELEX Collaboration), 
Phys. Rev. Lett. {\bf 89}, 112001 (2002).

A.~Ocherashvili {\it et al.\/} (SELEX Collaboration), 
Phys. Lett.~B {\bf 628}, 18 (2005).


\bibitem{CDF_SVT} J. Adelman {\it et al.\/} (CDF Collaboration), 
Nucl. Instrum. Meth. A {\bf 572}, 361 (2007).


\bibitem{EOI} The NuSOnG Expression of Interest is available from 
the Fermilab Directorate or at {\tt http://www-nusong.fnal.gov}.


\bibitem{NuTeVbeam}   R. Bernstein {\it et al.}, ``Sign-Selected Quadrupole Train'', 
FERMILAB-TM-1884 (1994).
J. Yu {\it et al.}, ``NuTeV SSQT Performance'', FERMILAB-TM-2040 (1998).


\bibitem{CCFR} 
W.\ G.\ Seligman {\it et al.} (CCFR Collaboration) 
Phys. Rev. Lett {\bf 79}, 1213 (1997).

U.\ K.\ Yang {\it et al.} (CCFR Collaboration)
Phys. Rev. Lett {\bf 86} 2742 (2001).


\bibitem{NuTeV} 
  M.~Tzanov {\it et al.}  (NuTeV Collaboration),
  Phys.\ Rev.\  D {\bf 74}, 012008 (2006).

\bibitem{CDHS}
  J.~P.~Berge {\it et al.},
  Z.\ Phys.\  C {\bf 49}, 187 (1991).
 
\bibitem{CHARMII}
 P. Vilain {\it et al.}, Phys. Lett. B {\bf 335}, 246 (1994).

\bibitem{NOMAD}  
  Q.~Wu {\it et al.}  (NOMAD Collaboration),
  Phys.\ Lett.\  B {\bf 660}, 19 (2008).

\bibitem{CHORUS}
  E.~Eskut {\it et al.}  (CHORUS Collaboration),
  Nucl.\ Phys.\  B {\bf 793}, 3 (2007).

\bibitem{SuperK}   C.~W.~Walter  (Super-K Collaboration),
  Nucl.\ Instrum.\ Meth.\  A {\bf 503}, 110 (2003).



\bibitem{K2K}   M.~H.~Ahn {\it et al.}  (K2K Collaboration),
  Phys.\ Rev.\  D {\bf 74}, 072003 (2006).


\bibitem{MINOSnumu}   P.~Adamson {\it et al.}  (MINOS Collaboration),
  Phys.\ Rev.\ Lett.\  {\bf 101}, 221804 (2008).


\bibitem{SNO}   B.~Aharnim {\it et al.}  (SNO Collaboration),
  Phys.\ Rev.\ C  {\bf 75}, 045502 (2007).

\bibitem{Bugey}
  B.~Achkar {\it et al.},
  Phys.\ Lett.\  B {\bf 374}, 243 (1996).

\bibitem{Chooz}
 M.~Apollonio {\it et al.}  (CHOOZ Collaboration),
  Phys.\ Lett.\  B {\bf 466}, 415 (1999).
  
\bibitem{MUNU}  Z.~Daraktchieva {\it et al.}  (MUNU Collaboration),
  Phys.\ Lett.\  B {\bf 615}, 153 (2005).


\bibitem{mudk}      
  C.~A.~Gagliardi, R.~E.~Tribble and N.~J.~Williams,
  Phys.\ Rev.\  D {\bf 72}, 073002 (2005). 

\bibitem{ahrens}  L.~A.~Ahrens {\it et al.},
  Phys.\ Rev.\  D {\bf 41}, 3297 (1990).  

\bibitem{'tHooft:1971ht}
  G.~'t Hooft,
  Phys.\ Lett.\  B {\bf 37}, 195 (1971).

\bibitem{lep2}
M.~S. Carena {\it et al.}, Phys.\ Rev.\ D {\bf 70}, 093009 (2004).

\bibitem{moller}
P.~L. Anthony {\it et al.}  (SLAC E158 Collaboration),
Phys.\ Rev.\ Lett.\  {\bf 95}, 081601 (2005).



\bibitem{Perl:1975bf}
  M.~L.~Perl {\it et al.},
  Phys.\ Rev.\ Lett.\  {\bf 35}, 1489 (1975).

\bibitem{Kodama:2008zz}
  K.~Kodama {\it et al.},
  Phys.\ Rev.\  D {\bf 78}, 052002 (2008).

\bibitem{Kodama:2000mp}
  K.~Kodama {\it et al.}  (DONUT Collaboration),
  Phys.\ Lett.\  B {\bf 504}, 218 (2001).



\bibitem{Abe:2006fu}
  K.~Abe {\it et al.}  (Super-Kamiokande Collaboration),
  Phys.\ Rev.\ Lett.\  {\bf 97}, 171801 (2006).
  
\bibitem{Fukuda:2000np}
  S.~Fukuda {\it et al.}  (Super-Kamiokande Collaboration),
  Phys.\ Rev.\ Lett.\  {\bf 85}, 3999 (2000).
  
\bibitem{ArgoNeuThttp}
 ArgoNeuT homepage: {\tt http://t962.fnal.gov}.
\bibitem{MicroBooNEprop}
 MicroBooNE collaboration, ``A Proposal for a New Experiment Using
 the Booster and NuMI Neutrino Beamlines: MicroBooNE'' (2007).

\bibitem{Amerio:2004ze}
  S.~Amerio {\it et al.}  (ICARUS Collaboration),
  Nucl.\ Instrum.\ Meth.\  A {\bf 527}, 329 (2004).

\bibitem{DUSELhttp}
  DUSEL homepage: {\tt http://www.lbl.gov/nsd/homestake}.

\bibitem{Amsler:2008zzb}
  C.~Amsler {\it et al.}  (Particle Data Group),
  Phys.\ Lett.\  B {\bf 667}, 1 (2008).

\bibitem{Itow:2001ee}
  Y.~Itow {\it et al.}  (The T2K Collaboration),
  arXiv:0106019 [hep-ex] (2001).

\bibitem{Badertscher:2004py}
  A.~Badertscher, M.~Laffranchi, A.~Meregaglia and A.~Rubbia,
  New J.\ Phys.\  {\bf 7}, 63 (2005).
  A.~Badertscher, M.~Laffranchi, A.~Meregaglia, A.~Muller and A.~Rubbia,
  Nucl.\ Instrum.\ Meth.\  A {\bf 555}, 294 (2005).
  A.~Ereditato and A.~Rubbia,
  Nucl.\ Phys.\ Proc.\ Suppl.\  {\bf 155}, 233 (2006).

\bibitem{Astier:2001yj}
  P.~Astier {\it et al.}  (NOMAD Collaboration),
  Nucl.\ Phys.\  B {\bf 611}, 3 (2001).

\bibitem{Davidson:2003ha}
  S.~Davidson, C.~Pena-Garay, N.~Rius and A.~Santamaria,
  JHEP {\bf 0303}, 011 (2003).
  
\bibitem{Friedland:2005vy}
  A.~Friedland and C.~Lunardini,
  Phys.\ Rev.\  D {\bf 72}, 053009 (2005).
  
\bibitem{Hirsch:2002uv}
  M.~Hirsch, E.~Nardi and D.~Restrepo,
  Phys.\ Rev.\  D {\bf 67}, 033005 (2003).
  






\bibitem{Asaka} T. Asaka and M. Shaposhnikov, Phys. Lett. B {\bf 620}, 17 (2005).
  M. Shaposhnikov, Nucl. Phys. B {\bf 763}, 49 (2007).
  M. Shaposhnikov, ``Dark Matter: The Case of Sterile Neutrino'', Proc. of Berlin 2006: Marcel Grossmann Meeting on General Relativity, Berlin, Germany (2007).
  A.~D. Dolgov {\it et al.}, Nucl. Phys. B  {\bf 590}, 562 (2000). M. Viel {\it et al.}, 
  Phys. Rev. D {\bf 71}, 063534 (2005).

\bibitem{Levy} J.~M. Levy, Ph.D. thesis, University of Paris, 1986 (unpublished).
 
\bibitem{Bernardi} G. Bernardi et al., Phys. Lett. B {\bf 166},  479 (1986).

\bibitem{minerva} 
  D. Drakoulakos {\it et al.\/} (Minerva Collaboration), arXiv:0405002 [hep-ex] (2004).

\bibitem{Annala} G. Annala, private communication.

\bibitem{Harding} D. Harding, Fermilab internal memo, 26 March 2008 (unpublished).

\bibitem{MTF} A. Mokhtarani, ``Results of Tevatron Dipole Tests at 3.6~K", MTF-93-0010 (1993).

\end{thebibliography}
\end{document}